%% file: TM09.tex
\documentclass[12pt,preprint]{aastex}
\usepackage{apjfonts}
\usepackage{amsmath}
\usepackage{epsfig}
\usepackage{graphicx,epsfig}
\usepackage{natbib}
\input{TM09_preamble.tex}

\shorttitle{Time-Dependence of MBH Merger Afterglows}
\shortauthors{Tanaka and Menou}

\begin{document}

\title{Time-Dependent Models for the Afterglows of Massive Black Hole Mergers}

\author{Takamitsu Tanaka and Kristen Menou}

\affil{Department of Astronomy, Columbia University, 550 West 120th Street, New York, NY 10027}

\begin{abstract}
The {\it Laser Interferometer Space Antenna} ({\it LISA})
will detect gravitational wave signals from
coalescing pairs of massive black holes in the total mass range
$(10^{5} - 10^{7})/(1+z)\;\Msol$ out to cosmological distances.
Identifying and monitoring the electromagnetic counterparts of these events
would enable cosmological studies and offer
new probes of gas physics
around well-characterized massive black holes.
\cite{MP05} proposed that
a circumbinary disk around a  binary of mass $\sim 10^{6}\Msol$
will emit an accretion-powered X-ray afterglow approximately one decade after the gravitational wave event.
We revisit this scenario by using Green's function solutions
to calculate the temporal viscous evolution 
and the corresponding electromagnetic signature
of the circumbinary disk.
Our calculations suggest that
an electromagnetic counterpart may become observable as a rapidly brightening
source soon after the merger, i.e. several years earlier than previously thought.
The afterglow can reach super-Eddington luminosities
without violating the local Eddington flux limit.
It is emitted in the soft X-ray by the innermost circumbinary disk,
but it may be partially reprocessed at optical and infrared frequencies.
We also find that the spreading disk becomes
increasingly geometrically thick close to the central object as it evolves,
indicating that the innermost flow could become advective and 
radiatively inefficient, and generate a powerful outflow.
We conclude that the mergers of massive black holes detected by {\it LISA}
offer unique opportunities for monitoring on humanly tractable timescales
the viscous evolution of accretion flows and the emergence of outflows
around massive black holes with precisely known masses, spins and orientations.
\end{abstract}

\section{Introduction}
\label{sec:intro}
Massive black holes (MBHs) are abundant at all observable redshifts,
manifesting themselves most often as active galactic nuclei (AGN) and quasars \citep{KR95, Richstone+98}.
A merger of a pair of galaxies each containing a MBH will result in the formation of a
MBH binary \citep{Begelman+80}.  Given the hierarchical structure formation history of the universe,
and that most or all formed galaxies appear to harbor a MBH in their nuclei \citep[e.g.,][]{Maggor+98},
there should be many MBH binaries formed throughout cosmic time
\citep{Haehnelt94, Menou+01, Volonteri+03, WL03, Sesana+07, TH09}.
Once compact, these binaries rapidly lose
orbital energy via gravitational radiation (GW) and coalesce.  The 
{\it Laser Interferometer Space Antenna} ({\it LISA}) will detect
gravitational waves resulting from MBH mergers of binaries with total mass
$M=(10^{5}-10^{7})/(1+z)\Msol$ out to redshift $z\sim 20$ \citep[e.g.,][]{Schutz09}.  {\it LISA} is expected
to be able to constrain the individual redshifted mass $(1+z)M$, the mass ratio $q\le1$,
spins of the MBHs, and
the luminosity distance of the source with precision, thereby
providing an unprecedented test of general relativity and probing the
assembly history and demography of MBHs
\citep[e.g.,][]{Hughes02, Vecchio04, LH06}.
However, the space-based detector will be unable to measure the redshift
(due to the fundamental degeneracy between source mass and redshift)
or the precise angular location 
(due to the fact that sources are located using the modulation of the signal
due to orbital motion and separation of the detector elements)
of the source 
\citep{Cutler98, HH05, Kocsis+06, Kocsis+07, LH08}.

The coalescence of a MBH binary is not by itself expected to emit an observable
electromagnetic (EM) signal.  If, however, MBH mergers have {\it associated}
and readily identifiable EM signatures, their detection would complement {\it LISA} by
helping determine source redshifts and locations on the sky \citep{HH05, Kocsis+06, Bloom+09, Phinney09}.
An obvious candidate to effect
such emission is the abundant gas linked to galactic MBHs.
Rapid, sustained gas accretion onto MBHs
is required to explain the luminosities and number of observed AGN and quasars.
Numerous studies suggest that a gas-rich environment aids
the formation of close MBH binaries in recently merged galaxies
\citep{Escala+05b, Dotti+07, Callegari+09, Colpi+09},
and may facilitate the further shrinking of the orbit to where GW emission
can enact the merger \citep{IPP99, AN02, MM08, Cuadra+09, Lodato+09}.
Various types of interactions between MBH
binaries and their gas environments have been proposed as viable mechanisms
for observable EM emission. 
Numerous studies
have calculated the EM emission
from the response of the gas disk to the mass loss and gravitational recoil \citep{Peres62, Bekenstein73}
effects
which accompany the merger process
\citep[e.g.,][]{BP07, SK08, Lippai+08, SB08, O'Neill+09, Megevand+09, Corrales+09, Rossi+09, Anderson+09}.
\cite{Chang+09} have suggested a luminous EM signal from
tidal and viscous excitation of fossil gas trapped inside the binary's orbit.
\cite{Krolik10} pointed out that even small amounts of gas
that is present in the immediate vicinity of a merging binary
can power short afterglows on Eddington luminosity scales.
\cite{LK08} have proposed an infrared afterglow
from the dissipation of GW through a surrounding gas disk
\citep[see, however, ][]{Krolik10}.
\cite{Palenz+09a, Palenz+09b, Mosta+09} have emphasized the possibility of variable
emission due to the perturbation and enhancement of the local EM fields.
\cite{HKM09} have raised the possibility of detecting
the binary as a periodic variable source prior to the merger.

In this paper we revisit the accretion afterglow mechanism
proposed by \citeauthor{MP05} (2005; hereafter MP05; see also \citealt{Liu+03, Liu04}) for a geometrically
thin circumbinary disk.  We summarize their model as follows:
\begin{enumerate}
\item The tidal torques from the binary open a gap
in the circumbinary gas.  The gas inside the gap accretes,
while the gas outside is held at bay by tidal torques.
What results is a circumbinary disk with the binary
inside the central cavity.  Because it is largely ``missing'' the inner, highest-temperature region,
the disk cannot easily produce thermal X-rays via viscous dissipation
(see, however, \citealt{Chang+09}).
The system can remain in this configuration for an extended period of time
\citep{IPP99, MM08}
as the timescale for the extraction of angular momentum from the binary's orbit is long.
\item Once the binary reaches
a semimajor axis $a\sim 100 GM/c^{2}$, it rapidly loses
orbital angular momentum and energy via GW emission and the orbit closes faster than the surrounding gas
can viscously follow \citep[see also][]{AN05, HKM09, Chang+09}.
The binary coalesces, producing the GW signature detectable by {\it LISA}.
\item The gas (no longer held back by binary torques) reaches the center of the cavity
by viscously spreading, and the corresponding accretion flow,
deep in the potential well of the MBH remnant, emits an X-ray afterglow.
\end{enumerate}
By taking the difference between the time it takes the binary to merge
and the time it takes for the {\em bulk} of the gas at the cavity edge to reach the central remnant,
\citetalias{MP05} estimated that the X-ray afterglow would occur
$t_{\rm EM}\sim 7(M/10^{6}\Msol)^{1.3}(1+z)$ yr after the GW signal (in the observer's rest frame).

In this study, we use the term ``accretion afterglow'' to denote
the emission described above.  
We stress that this is a distinct mechanism from the various
afterglow mechanisms that are powered by the response of the accretion
disk to the mass loss and/or recoil of the central MBH remnant (references above).
In the latter scenarios, 
mass loss and/or recoil introduce additional eccentricities in the gas orbits,
and the emission is powered by the circularization of the orbits and shock/wave dissipation.
In the accretion afterglow scenario considered by \citetalias{MP05} and
in this paper, the emission results from the deepest parts of
the MBH potential well becoming accessible to the circumbinary disk as the binary shrinks and merges.
Before the merger the disk is deficient
in the hard UV and X-ray frequency range because of the central cavity,
while after the merger this high-frequency emission emerges as the cavity fills with gas.
Although distinct, the various classes of emission mechanisms considered are not entirely unrelated.
For example, mass loss introduces a roughly constant eccentricity
everywhere in the disk, and the resulting luminosity of
a circularization-powered afterglow (references above)
would depend delicately on the innermost density profile
of the circumbinary disk at the time of merger (see \citealt{Corrales+09} for a comparison
of afterglows for different density profiles).
A high surface density for the innermost disk (where the
available specific orbital energy is highest) at merger would enhance the mechanism
emphasized by \cite{O'Neill+09} while it would suppress the pre- and post-merger emission
contrast considered in this paper; a low surface density
would have the reverse consequences for the two mechanisms.

The purpose of this work is to investigate the onset and time dependence
of the afterglow in greater detail.
We use an idealized semi-analytic framework
to model the time dependence of the electromagnetic signal.
We present explicit, integral solutions for the time evolution of
one-dimensional (viz. geometrically thin and azimuthally symmetric), viscous Keplerian disks.
(See similar work by \citealt{Shapiro10}.)
Our calculation method is valid insofar as the disk kinematic viscosity
can be described as a simple function of radius,
and as long as after decoupling the inner gas is minimally
affected by the gravitational torques from the binary.
This last point is worth emphasizing, as in general
the decoupling of the binary from the gaseous influence does not
guarantee that the gas is entirely free from the binary influence.
Even after GW emission has
become the dominant mechanism driving the orbital evolution
of the binary, the binary's tidal torques can still influence
any gas that is able to remain in the vicinity of the binary's orbit.

This paper is organized as follows.
In \S \ref{sec:model} we review the disk properties as
the GW-driven closing of the binary decouples from
the viscosity-driven spreading of the circumbinary disk,
and introduce a semi-analytic model for the subsequent evolution
of the disk's surface density.  
Various derivations and intermediate results used in this section
are provided in the Appendices \ref{sec:AppA} \& \ref{sec:AppB}.
We present in \S \ref{sec:obs} the predictions of the model
for the light curve and spectral evolution of the resulting electromagnetic emission.
We also discuss in that section the possibility that the X-ray
afterglow could be reprocessed by the outer disk, as well as
possible effects of advection as the inner disk becomes geometrically thick.
We conclude  in \S \ref{sec:concl}.

\section{Modeling the Binary-Disk System}
\label{sec:model}
Below, we first discuss the properties introduced by \citetalias{MP05}
for the evacuated circumbinary gas disk at the time when
the evolution of the binary orbit becomes dominated by GW emission
and thus decoupled from the evolution of the surrounding disk,
which evolves viscously (see also \citealt{Haiman+09}).  We then describe the semianalytic
integral formalism for the viscous evolution of an accretion disk with an
arbitrary initial surface density profile.  Additional derivations and intermediate results
for both the disk properties and the semianalytic model are 
detailed in Appendices A and B, respectively.
Throughout this paper, $R$ is the radial distance from the center of mass
of the binary in the plane of its orbit; $a$ is the binary's semimajor axis;
$M$ is the total mass of the binary; and $q\le 1$ is the binary mass ratio.
The quantities
$c$, $G$, $m_{p}$, $h$, $k$, $\sigma $ and $\sigma_{\rm T}$ are
the speed of light, the gravitational constant, the mass
of the proton, Planck's constant, the Boltzmann constant,
the Stefan-Boltzmann constant and the Thomson cross section, respectively.
Times, frequencies and rates are in the rest frame of the
binary, unless noted otherwise.

\subsection{The Circumbinary Disk at Decoupling}
The ``cold'' gas in the nucleus is assumed to settle into a 
geometrically thin, rotationally supported disk.  
In this paper, we are concerned with the properties of the innermost
disk and the binary in the very last stages of its evolution.
We assume that the angular momentum of the
disk is aligned with the orbital angular momentum
of the binary (i.e., no warps; \citealt{IPP99}; however we shall
revisit the possibility of warping later), and that the binary
has been circularized by interactions with the surrounding gas (see 
\citealt{AN05}, \citealt{MM08} and \citealt{Cuadra+09} for possible complications).

The disk is able to evolve locally on length scales $\sim R$
on the viscous timescale
\beq
t_{\rm visc}(R)=\frac{2}{3}\frac{R^{2}}{\nu},
\eeq
where $\nu(R)=\frac{2}{3}\alpha P_{\rm gas}/(\rho\Omega)=\frac{2}{3}\alpha kT/(\mu m_{\rm p}\Omega)$
and $P_{\rm gas}, \rho$, $T$ and $\mu=0.6$ are the pressure,
density, temperature and mean molecular weight of the gas, respectively.
The quantity $\Omega$ is the angular velocity, assumed to be approximately Keplerian:
$\Omega\approx \sqrt{GM/R^{3}}$.
The parameter $\alpha$ is the viscosity parameter in the disk model of \citep{SS73}.
We follow \citetalias{MP05} in adopting a
viscosity prescription where $\nu$ scales only with the gas pressure,
as opposed to the total (gas, radiation plus magnetic) pressure.
This choice is motivated by calculations that suggest that
disks where viscosity scales with the total pressure
may be thermally unstable (\citealt{SS76, Pringle76}; see, however, \citealt{HKB09}).
\footnote{
However, radiation pressure dominated disks may still be viscously unstable
\citep{LE74, Piran78, HBK09}.
Because the afterglow mechanism discussed in this paper
takes place in the radiation-dominated part of the disk,
the findings reported here are contingent on the answer to this open theoretical question.
}

At a distance of $R\sim 2a$, a gap is opened in the circumbinary disk
through tidal interactions (see, e.g., \citealt{Artymo+91}, \citealt{MM08} and \citealt{Haiman+09}).
In this region the input of angular momentum from the
 tidal torques exerted by the binary counteracts the
viscous spreading of the disk.
For simplicity, we neglect the small amount
of fossil gas that may be present inside the binary's orbit \citep{Chang+09}.
The gap, at first annular, becomes a roughly circular central cavity after the inner gas is consumed.
The boundary between the cavity and the disk is
characterized by a steep surface density gradient,
a pile-up of gas caused by
tidal {\it decretion} in the vicinity of the binary's orbit
and  viscous accretion outside the decreting region.
Whereas a steady-state accretion disk has a mass accretion rate
$\dot{M}(R)=3\pi\nu\Sigma$ that is constant with radius,
a decretion disk with a steady inner torque instead satisfies constant viscous torque,
i.e. $3\pi R^{2}\nu\Sigma\Omega=$ constant \citep[e.g.,][]{RP81, Pringle91}.
It is a priori unclear how close to accretion or decretion a certain system is at decoupling,
as this will depend on the system parameters and its past accretion history
\citep[e.g.,][]{IPP99, Chang+09}.
We thus consider various possibilities for the circumbinary disk density profile at decoupling.
Some of the innermost gas may be still able to accrete across the binary's orbit
in narrow streams, but the accretion rate is expected to be only a small fraction of that of
a comparable accretion disk around a single black hole
\citep[e.g.,][]{LSA99, Gunther+04, Hayasaki+07, Hayasaki+08, MM08}.

The timescale for the binary separation to shrink via GW emission is
\beq
t_{\rm GW}(a)\equiv\frac{a}{da/dt}=\frac{5}{16}\frac{c^{5}}{G^{3}M^{3}}
\frac{a^{4}}{\zeta}\approx 4.90\yr \times \left(\frac{a}{100 GM/c^{2}}\right)^{4}\zeta^{-1},
\label{eq:tGW} 
\eeq
where $\zeta\equiv 4q/(1+q)^{2}$ is the symmetric mass ratio scaled to unity
for equal-mass binaries.  Equation \eqref{eq:tGW} is valid for binaries with non-extreme
mass ratios and circular orbits.  Because $t_{\rm GW}\gg t_{\rm visc}$ initially
in the circumbinary disk, the gas is able to respond promptly
to the gradual and relatively slow shrinking of the binary, probably maintaining
a roughly constant geometric ratio between the disk inner edge and the
orbital semimajor axis, i.e. $\lambda\equiv R/(2a) \sim 1$.

Because the disk viscosity $\nu$ is a weak function of radius outside
the inner edge (see Appendix \ref{sec:nval}),
roughly speaking the viscous time there scales as
$t_{\rm visc}(R)\propto R^{2}$.  As the binary closes,
$t_{\rm GW}(a)\propto a^{4}$ will inevitably become shorter than
the disk viscous time at the inner edge, and the binary will begin to close faster
than the bulk of the gas at the inner edge can follow.
The subsequent evolution of the binary is driven by GW emission
and thus causally decoupled from that of the disk.
In the absence of binary torques, the gas at the cavity edge
would fill the cavity in a time $\sim \beta t_{\rm visc}$, where
the approximate boost factor $\beta\sim 0.1$
was introduced by \citetalias{MP05} to account for the limit
of a very steep surface density gradient at the edge \citep{LP74}.
Decoupling occurs, then, when $t_{\rm GW}(a)\sim \beta t_{\rm visc}(2\lambda a)$.
This condition, along with the assumption that the viscously dissipated energy
is locally emitted as thermal radiation, specifies the various properties of the inner edge
of the disk described by \citetalias{MP05}.
We closely reproduce their results in Table \ref{tab:1}, and detail the various
intermediate calculations in Appendix \ref{sec:AppA}.
Note that while we recover the parameter dependencies exactly,
the magnitudes of our disk properties differ somewhat from those of \citetalias{MP05}.
This is due to differences in how we have calculated the thermal structure of the disk,
the most significant being that the flux expression in \citetalias{MP05}
is a factor of two lower than ours (compare their equation [2] and preceding formula for $F_{\nu}$
with our equation [\eqref{eq:modF1}] below).
Throughout the rest of this paper, we use the subscript ``0'' to denote 
the value of a quantity at decoupling.
When applied to a disk quantity with radial dependence, the subscript
shall also denote the value at the inner edge of the disk.
For example, $a_{0}$ is the semimajor axis of the binary orbit at decoupling
and $\Omega_{0}$ is the angular velocity at the inner edge of the disk at decoupling.

The surface density at the inner edge of the circumbinary disk at decoupling cannot be
determined through scaling arguments alone.
However, we expect it to be greater than the value expected for a steady thin accretion disk.
Just outside the edge, the gas may resemble a decretion disk, with
$3\pi\nu\Sigma\propto R^{-1/2}$ approximately, because of mass accumulation.
Sufficiently far from the binary and its tidal torques, we expect the gas to behave more like a steady thin accretion disk,
with a local mass supply rate $\dot{M}=3\pi\nu\Sigma$ roughly constant and determined by
external conditions \citep[e.g.,][]{IPP99, Chang+09}.
It follows that if $3\pi\nu\Sigma$ is monotonic outside the edge
then it must be greater than the outer mass supply rate of the disk.
Therefore, we parametrize the surface density at the inner edge via the arbitrary
relation $\Sigma_{0}=S\dot{M}_{\rm Edd}/(3\pi\nu_{0})$, where 
$S$ is a dimensionless parameter and 
$\dot{M}_{\rm Edd}=4\pi GMm_{\rm p}/(\eta c\sigma_{\rm T})$
is the Eddington accretion rate onto a MBH of mass $M$ for a 
radiative efficiency $\eta=0.1$.
Effectively, $S$ is the product of the gas supply rate onto the disk
in Eddington units and the enhancement of the local surface density at the disk edge
due to the binary torques and mass accumulation.\footnote{
For practical purposes $S$ is equivalent to the parameter ``$\dot{m}$'' used in \citetalias{MP05}.
We avoid using the latter notation to prevent confusion with
the actual local accretion rate, which is discussed below in \S \ref{sec:viscev}.
Also note that $\Sigma_{0}$ and $S$ do not relate linearly
(see Table \ref{tab:1}).  The relationship is more complex because the disk gas viscosity
$\nu_{0}$ and the decoupling orbital radius $a_{0}$ also depend on $S$
through the midplane temperature.
}
In general, $S$ can exceed unity even if the disk mass supply rate
and the local accretion rate are both sub-Eddington.

\subsection{A Simple Model for the Viscous Evolution}
\label{sec:viscev}
Following decoupling, the binary outruns the surrounding gas to
the center and merges in a time $t_{\rm GW,0}/4$.
\citetalias{MP05} estimated that an accretion-powered X-ray
afterglow would take place when the central cavity is filled, viz.
after a time
$t_{\rm EM}= \beta t_{\rm visc, 0}-t_{\rm GW,0}/4 \sim (3/4)t_{\rm GW,0}$ after the GW event.
Below, we revisit this estimate by calculating
the time dependence of the signal with a simple model
for the disk evolution.  We proceed by assuming that
following decoupling the binary's orbit closes faster than the
gas can follow, viz. viscosity is the only source of
torque in the disk and the gas orbits remain circular and Keplerian.
Numerical simulations \citep[e.g.,][]{AN05, Hayasaki+07, MM08, Cuadra+09}
of MBH circumbinary disks 
indicate that both the eccentricity and the
deviation of the angular velocity from the Keplerian value
are small (of order a few percent) at the radius where the binary tides
truncate the disk.
Likewise, the mass reduction of the binary
at merger due to GW emission will only introduce
orbital eccentricities (roughly equal to the
fractional mass loss) of $\gtrsim 10^{-2}$ \citep{TM08, O'Neill+09, Reisswig+09}.
At radii and binary masses of interest here, orbital speeds are too great
for the gas to be significantly affected by gravitational recoil of the MBH remnant:
$\Omega\sim \sqrt{Rc^{2}/GM}c\gg v_{\rm recoil}\sim 300 \km \s^{-1}$.
After the binary has merged,
the central potential is that of a rotating black hole
and circular orbits near the center will have somewhat super-Keplerian
angular velocities.
Approximate ``pseudo-Newtonian'' expressions of the Kerr potential
\citep[e.g.,][]{Artemova+96} suggest that for a merger
remnant with less than maximal spin, the deviation from
the Keplerian value is of order ten percent at the radius
of innermost stable circular orbit and falls off as roughly $\propto R^{-1}$.
All of these modest deviations and perturbations are neglected in our models.

The viscous evolution of a geometrically thin, cylindrically
symmetric Keplerian disk can be described by the standard equation
 (e.g. see \citealt{Pringle81} and \citealt{FKR02})
\beq
2\pi R\frac{\dd}{\dd t}\Sigma (R,t)=\frac{\dd}{\dd R}\left[2 R^{1/2}\frac{\dd}{\dd R}\left(3\pi\nu\Sigma R^{1/2}\right)\right],
\label{eq:diff}
\eeq
where $\Sigma$ is the surface density of the disk and the innermost
derivative on the right-hand side describes the angular momentum gradient.
The left-hand side of \eqref{eq:diff} describes the rate of change of the radial mass distribution,
while the right-hand side gives the radial gradient of the local mass flow
\beq
\dot{M}=3\pi\nu\Sigma\left(1+2m\right),
\label{eq:Mdot}
\eeq
where $m(r)\equiv \dd \ln (\nu\Sigma)/\dd \ln R$.
In the above equation,
the dimensionless factor in parentheses
equals unity for steady-state accretion solutions ($m\approx 0$)
while it vanishes in steadily decreting solutions ($m\approx -1/2$).

If the kinematic viscosity is a function of radius only, equation \eqref{eq:diff} is a linear diffusion equation.
In the special case $\nu\propto R^{n}$, the solution for subsequent viscous evolution
(see, e.g. \citealt{LP74}; a derivation is provided in our Appendix \ref{sec:AppB}) is given by
\beq
\Sigma(r,t)=\int_{0}^{\infty}G(r,r^{\prime},t)\Sigma(r^{\prime},t=0)dr^{\prime},
\label{eq:sigsol}
\eeq
where 
\begin{eqnarray}
G(r,r^{\prime},t)&=&\frac{2-n}{\tau}r^{-1/4-n}r^{\prime 5/4}
I_{1/(4-2n)}\left(\frac{2r^{\prime 1-n/2}r^{1-n/2}}{\tau}\right)
\nonumber\\
&&\qquad \times 
\exp\left(-\frac{r^{\prime 2-n}+r^{2-n}}{\tau}\right)
\left\{1-\sqrt{\frac{r_{*}}{r}}
\exp\left[-\frac{\left(r-r_{*}\right)^{2-n}}{{\mathcal R}(\tau-\tau_{\rm merge})}\right] \right\}
\label{eq:Green}
\end{eqnarray}
is the Green function, $\Sigma(r,t=0)$ is the initial surface density profile (at decoupling),
$I_{m}$ is the modified Bessel function of the first kind,
${\mathcal R}$ is the ramp function,
and $r\equiv R/R_{0}$, $r^{\prime}\equiv R^{\prime}/R_{0}$
and $\tau\equiv 8(1-n/2)^{2}t/t_{\rm visc,0}$ are dimensionless variables.
At decoupling, $\tau=0$; the binary merges when $\tau=\tau_{\rm merge}=2\beta (1-n/2)^{2}$;
roughly speaking, gas concentrated at $r=1$ diffuses to the center in a time $\tau\sim 1$
(hence the value $\beta\sim 0.1$ adopted by \citetalias{MP05}).

The second term in the curled brackets  of
equation \eqref{eq:Green} accounts for the fact
that, after merger, the MBH remnant imposes a zero-torque
boundary condition at some finite radius $R_{*}=R_{0}r_{*}$,
which we associate with the remnant's innermost radius for
marginally stable circular orbits \footnote
{See, however, \cite{KH02}, who showed
that there are several ways to define an effective inner radius, some of
which are quite different from $R_{\rm ISCO}$.} ($R_{\rm ISCO}$).
This radius depends on the MBH spin as
$a_{\rm spin}=(1/3)\sqrt{R_{\rm ISCO}c^{2}/GM} (4 - \sqrt{3 R_{\rm ISCO}c^{2}/GM - 2})$ \citep{Bardeen70},
where $-1\le a_{\rm spin}\le 1$ is the standard dimensionless spin parameter of the MBH.
For test particle orbits, one finds $R_{*}=6GM/c^{2}$ if $a_{\rm spin}=0$,
and $R_{*}=GM/c^{2}$ if $a_{\rm spin}=1$ and the spin of the hole
is perfectly aligned with the test particle's orbital angular momentum.
Immediately after the merger, the MBH remnant
is expected to have moderately high but non-maximal
spin for a wide range of physically plausible scenarios
(e.g. \citealt{HB03, Baker+04, BV08}; see however the ``chaotic'' accretion scenario of
\citealt{KP06}, which predicts lower MBH spins).
We thus adopt the intermediate value $R_{*}=3GM/c^{2}$,
which is accurate to within $\sim 20\%$
in the range of spin $0.65\ltsim a_{\rm spin}\ltsim 0.9$.
Our solutions given by equation \eqref{eq:Green} depend
only on the initial surface density profile
(which may be arbitrarily complex and need not be differentiable),
as well as on the values of $n\le 2$ and $r_{*}$.

\begin{figure}
\centerline{\hbox{
\plotone{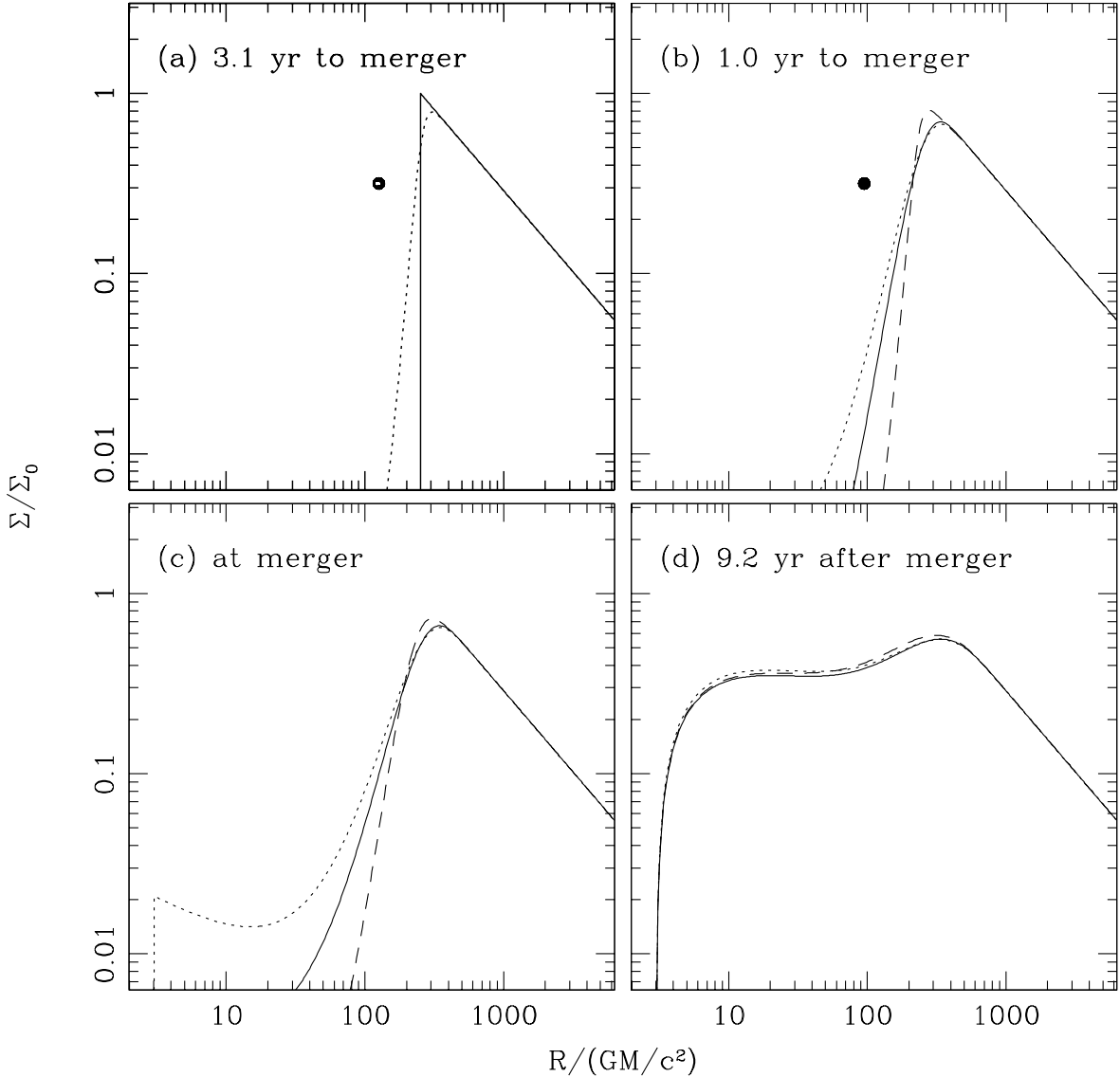}
}}
\caption{Snapshots of the time-dependent solutions to equation \eqref{eq:sigsol}
for the surface density $\Sigma/\Sigma_{0}$ for fiducial disk parameters.
We evolve the solution by applying \eqref{eq:Green} with $n=0.4$,
which is roughly consistent with the physical prescription 
$\nu\propto T/\Omega$ at late times and small radii, when the disk
approaches a quasi-steady accretion track satisfying $\Sigma\nu \propto (1-\sqrt{R_{*}/R})$.
Times in the figure are scaled to the merger of an equal-mass, $10^{6}\Msol$ binary.
Panel (a) shows the initial surface density profiles at decoupling.
Panels (b), (c) and (d) show the evolved profiles at 1 year before the merger, at merger, and
9.2 years after the merger, respectively.  We also evolve a third profile (dashed lines)
that qualitatively describes a scenario where the binary continues to open a gap even after
the nominal decoupling condition (see text for details).
In panels (a) and (b), the semimajor axis of the binary is shown schematically with black circles
(the orbit shrinks by $\approx 25\%$ between the panels).
After the merger, we impose a zero-torque condition at $R_{*}=3GM/c^{2}$;
surface density profiles at merger (Panel c) are truncated at this radius and subsequently evolved with
the new boundary condition.
All three profiles have the same qualitative time-dependent behavior, suggesting that
neither the initial surface density profile at decoupling nor the precise time
when the binary ceases to influence the disk edge are critical for our main conclusions on
the afterglow signatures.
}
\label{fig:sig1}
\end{figure}

Figure \ref{fig:sig1} shows time-dependent solutions 
calculated following equation \eqref{eq:sigsol} for
two different initial surface density profiles at decoupling (upper left panel).
In solid lines, we evolve for demonstrative purposes
the plainest possible model: a surface density profile
obeying the decretion power-law $\Sigma\propto r^{-1/2-n}$,
 truncated by a step function at the cavity edge.
We also evolve a second, more plausible profile (dotted lines):
\beq
\frac{\Sigma(r,0)}{\Sigma_{0}}=\frac{1}{r^{n+1/2}+\exp[-11.9\lambda(r-1)]}.
\label{eq:MM08}
\eeq
This analytic profile is inspired by the 
hydrodynamic simulation results of \cite{MM08},
who found an azimuthally averaged inner surface density profile with
$\Sigma (R\ltsim 2a)\propto  \exp(-5.95R/a)$
 for a thin disk around an equal-mass binary.
Equation \eqref{eq:MM08} reproduces this exponential behavior
for $r\ltsim 1$ and the decretion power-law behavior $\Sigma\approx \Sigma_{0} r^{-1/2-n}$ for $r\gtrsim 1$.
Note that this profile does not peak at the inner edge $r=1$, nor is its peak value equal
to the parametrized quantity $\Sigma_{0}$.  In the context of 
equation \eqref{eq:MM08}, the value of $\Sigma_{0}$ sets
the scale for the power-law region of the surface density profile just outside the inner edge.

\begin{figure}
\centerline{\hbox{
\plotone{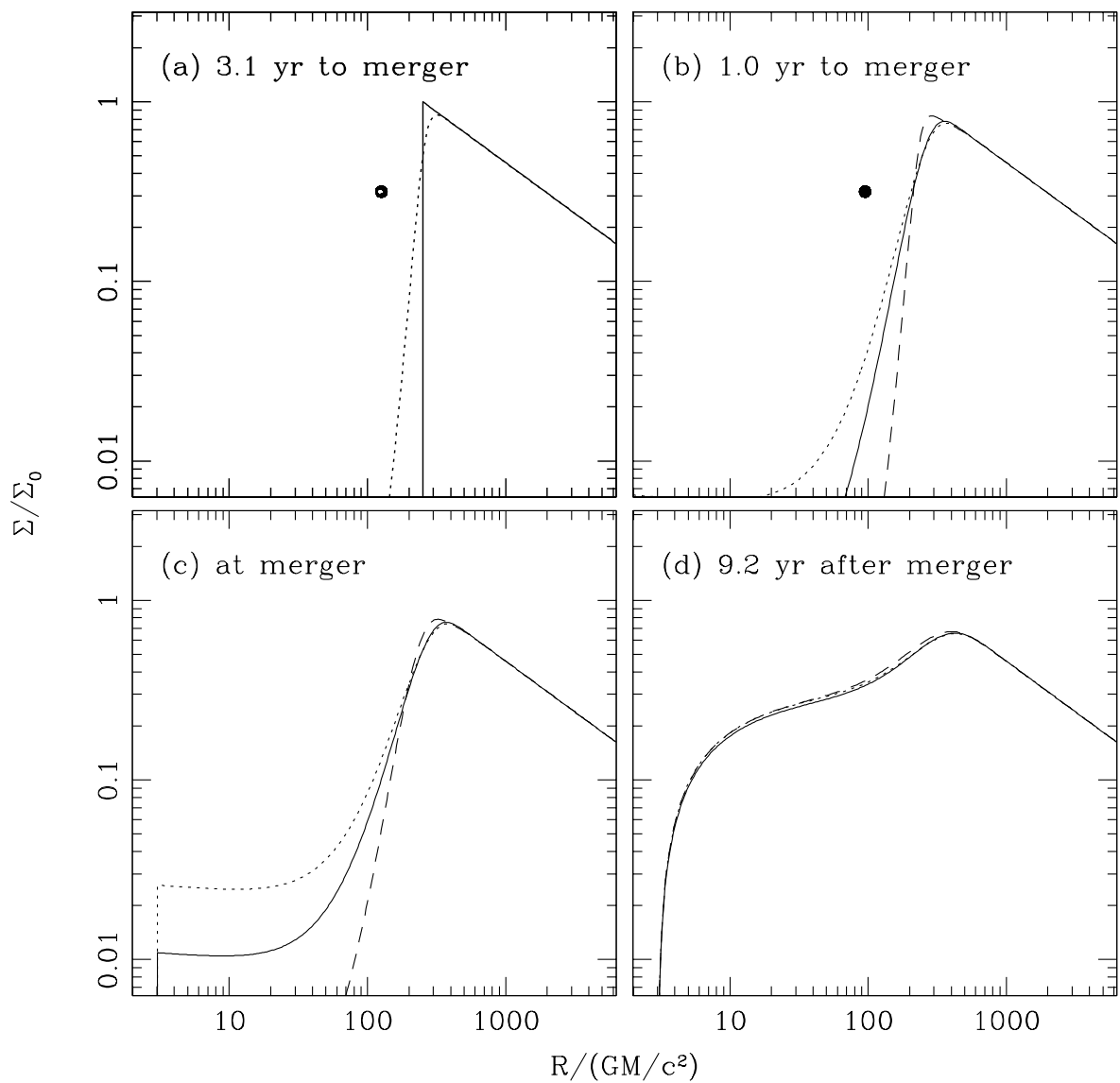}
}}
\caption{Same as Figure \ref{fig:sig1}, but with $n=11/170\approx 0.065$, which is the value
consistent with a physical prescription for viscosity, $\nu\propto T/\Omega$,
just outside the inner edge of the disk.  We find that our
qualitative findings are insensitive to the value of $n$, as long as the quantity
$2-n$ is not much smaller than unity.
}
\label{fig:sig2}
\end{figure}

In Figure \ref{fig:sig1}, the physical length scale and the time with respect
to the binary merger is calculated using the relations in Table \ref{tab:1} and
the fiducial parameter values $M_{6}=q=\alpha_{-1}=\beta_{-1}\lambda=s=\theta_{0.2}=1$.
We take as $n=0.4$ as our fiducial viscosity power-law.
For our disk solutions, this value leads to agreement between
the late-time disk profile $\nu\Sigma\propto 1-\sqrt{R_{*}/R}$ and
the physical viscosity prescription $\nu\propto P_{\rm gas}/(\rho\Omega)$
of the $\alpha_{\rm gas}$ disk.
(We refer the reader to Appendix \ref{sec:nval} for a brief discussion on
the parameter $n$.)
In Figure \ref{fig:sig2}, we also evolve a second case with $n=11/170\approx 0.065$, which is consistent
with $\nu\propto P_{\rm gas}/(\rho\Omega)$ at early times 
just outside the inner edge for a disk that has the decretion profile $m\approx -1/2$ there.
In both figures, we graph the profiles at decoupling (3.1 years before
the merger), 1 year before merger, at merger, and 9.2 years after merger.
The last snapshot corresponds to a time $\beta t_{\rm visc,0}$ after decoupling,
our re-evaluation of \citetalias{MP05}'s nominal estimate for the onset of the X-ray afterglow.

These two figures suggest that the qualitative evolution of
the disk does not depend very sensitively on $n$.
The viscous evolution timescale
governing the Green's function
scales as $(1-n/2)^{2}$, and thus the temporal dependence
on $n$ is small as long as the quantity $1-n/2$ is of order unity.
In both figures, the qualitative evolution is such that at early times
and large radii ($t<t_{\rm visc}(R)$), the disk maintains the initial
mass distribution.
At late times and small radii ($t\gtrsim t_{\rm visc}(R)$),
the disk approaches an analytic quasi-steady density profile.  Before the merger,
we do not specify an inner boundary condition and so the
quasi-steady surface density profile satisfies $\Sigma \propto \nu^{-1}\propto R^{-n}$.
After the merger, however, we assume that the MBH remnant imposes a zero-torque
boundary condition at finite radius $R_{*}$, which leads to a quasi-steady
surface density profile satisfying $\Sigma \propto (1-\sqrt{R_{*}/R})R^{-n}$.
The two initial profiles evolve to become largely indistinguishable
even before the binary merges.
For a variety of initial profiles, we have confirmed that the solutions converge on timescales
shorter than the time to merger if the
profiles share the same value of $\lambda$,
which is closely related to the location of the maximum of the surface density gradient,
and have identical profiles just outside the inner edge.

We note that the physical circumbinary disk is not likely maintain a $m\approx -1/2$
decretion profile
at radii far outside of the inner edge.  Where the gas is causally
detached from the binary torques and the mass accumulation
of the inner edge, it is expected to have the $m\approx 0$ profile
of a steady accretion disk.
The simulations of \cite{MM08} suggest that this transition to
a steady accreting solution does not occur until the radius
is at least several times the size of the central cavity.
Because the viscous time scales as $t_{\rm visc}(R)\propto R^{2-n}$,
we do not expect the gas profile at decoupling at these outer radii
to contribute to the EM afterglow on timescales of observational interest.

Despite decoupling, some gas is able to follow the binary closely.
This raises the possibility that some gas could continue to partially influence the subsequent orbital
evolution of the binary.  Such a dynamical effect will be proportional to the gaseous mass present in the vicinity
of the binary.  Since the surface density profile at decoupling cannot viscously evolve significantly
at radii beyond several times $R_{0}$, we choose to conservatively evaluate the maximum dynamical
influence of the circumbinary gas based on the mass inside $10R_{0}$.
All of the disk profiles in Figures \ref{fig:sig1} and \ref{fig:sig2}
have a total mass inside $R<10R_{0}$ of
\beq
M_{\rm disk}(R<10 R_{0})\sim 10^{-5}M\times
\alpha_{-1}^{-1.36}S^{0.02}\lambda^{2.8}M_{6}^{1.32}(\beta\zeta)^{0.7}\theta^{-0.34},
\label{eq:Mdisk}
\eeq
where the parameter dependencies above are calculated from
Table \ref{tab:1} and Appendix \ref{sec:AppA} using the
fact that $M_{\rm disk}\propto \Sigma_{0}\lambda^{2}a_{0}^{2}$.
The low, conservative value of the disk to binary mass ratio in
equation \eqref{eq:Mdisk} suggests that 
the disk is unlikely to have a dynamical effect on the binary merger,
and to contaminate the GW signal at levels detectable by {\it LISA}.
(Note, however, the scenario of \citealt{AN05}, in which the disk
can imprint {\it LISA}-observable perturbations on the binary's orbit prior to the decoupling time.)
However, such gas may still contribute an observable EM signature
because the available specific gravitational energy is large \citep{Chang+09, Krolik10}.

Even if the gas cannot influence the binary, it is possible for the binary to continue to influence
the evolution of the innermost gaseous disk, even after decoupling.
Our models assume that the evolution of the gas and that of the binary can
effectively be treated independently, i.e. that the region where the potential is non-Keplerian
shrinks faster than the gas can follow.
Although this is a useful first approximation,
it is not strictly guaranteed by a decoupling criterion
based on GW and viscous timescales, as defined earlier.
Indeed, our solutions allow for gas as dense as $\gtrsim 10^{-2}\Sigma_{0}$
to flow to $R\ltsim a(t)$ prior to the merger (see, e.g., Figure \ref{fig:sig1}b).  
To understand why this happens, one can compare the radial gas velocity
$V_{R}=-\dot{M}/(2\pi R\Sigma)=3(1+2m)\nu/(2R)$ with the
shrinking rate of the hypothetical disk edge, $d(2\lambda a)/dt$.
The decoupling condition $a_{0}/(da/dt)_{0}=\beta t_{\rm visc,0}$
directly implies $d(2\lambda a)/dt=\beta^{-1} R_{0}/t_{\rm visc,0}=3\beta^{-1}\nu_{0}/(2R_{0})$.
We thus find that the nominal radial velocity (ignoring binary torques)
at the inner edge at decoupling is given by the following expression:
\beq
V_{R,0}= 0.1\beta_{-1}\left(1+2m_{0}\right)\frac{d(2\lambda a)}{dt}.
\label{eq:VR}
\eeq
Above, $m_{0}$ is the index $\dd\ln(\nu\Sigma)/\dd\ln R$ evaluated at the disk edge,
which in principle may be much greater than unity.
For sufficiently steep surface density profiles,
the innermost gas can continue to flow into the binary's
vicinity\footnote{In fact, because our viscosity prescription is not causal, $V_{R}$ 
can be supersonic for large positive values of $m_{0}$.
Flow speeds are, however, generally subsonic for 
realistic $\Sigma$ profiles, as long as $\dd \ln \Sigma /\dd \ln R \ltsim 10^{3}$ at decoupling.
Note that the step-function profile shown in Figure \ref{fig:sig1}(a) rapidly evolves to a profile with subsonic
radial speeds throughout.} even after decoupling as defined earlier.
Thus, the binary can continue to open a gap in the disk for some time
until the binary is truly able to outrun the surrounding gas.
This happens when the binary's orbital radius has shrunk below the decoupling value $a_{0}$
by a factor $[0.1\beta_{-1}(1+2m)]^{1/(n+2)}$.
The profiles given by the analytic form in equation \eqref{eq:MM08} have $m_{0}\gtrsim 10$
for each of the viscosity prescriptions shown in Figures  \ref{fig:sig1}
and \ref{fig:sig2}.  We thus estimate the binary can outrun the gas
in these profiles
when $a\ltsim 0.8 a_{0} \beta_{-1}^{-1/(n+2)}$,
at a time approximately $1 M_{6}^{1.3} \yr$ before the merger.
This suggests that the time when the binary torques
cease to influence the circumbinary gas may be as late as $\sim 0.05 t_{\rm visc,0}$
before the merger.

To evaluate the possible effect of the binary's continued influence on the gas
after formal decoupling,
we consider a scenario where the inner gas has maintained an exponential
inner density profile similar to the one in equation \eqref{eq:MM08}, scaled
to the new semimajor axis of the binary, until one year before the merger
as estimated above.
This revised configuration is shown as dashed lines in Figures \ref{fig:sig1}
and \ref{fig:sig2}.  It is evolved viscously in the same way as the other two scenarios
in panels (c) and (d) of each figure.
While the first two classes of profiles represent the assumption that the binary
ceased to tidally interact with the circumbinary gas at decoupling
($t_{\rm GW}=\beta t_{\rm visc}$), the third class of profiles is evolved
assuming the tidal interaction continues for a while longer
after decoupling, until the binary is able to truly outrun
the disk edge as estimated above. 
We find that the three types of profiles are unlikely to be
observationally distinguishable from each other
(compare long-dashed and dotted lines in Panels b, c, d in Figures \ref{fig:sig1} and \ref{fig:sig2});
there are small differences until merger time, but they are quickly
wiped clean by the boundary condition associated with the single MBH remnant.
Figures \ref{fig:sig1} and \ref{fig:sig2} illustrate that
neither the density gradient at the inner edge at decoupling
nor the subsequent gas-binary interaction inside this radius
are likely to have an appreciable effect on the disk evolution
on timescales of interest.
At the level of accuracy of our idealized models,
we do not expect either of these factors to affect too strongly
the observable properties of the viscously spreading disk, to which problem we now turn.

\section{Observable Features of the Time-Dependent Afterglow}
\label{sec:obs}
\subsection{Bolometric Light Curve}
The power per unit area viscously dissipated from
each face of a thin Keplerian disk is
equal to $F\sim (9/8)\nu\Sigma\Omega^{2}$.
Substituting our expressions for $\nu\propto r^{n}$ and $\Omega\propto r^{-3/2}$, and 
ignoring advective loss, the power radiated by the disk is
\begin{eqnarray}
L_{\rm visc}(t)&\approx& \frac{9}{4}\nu_{0}\Omega_{0}^{2} \int  \Sigma(r,t)\;2\pi r^{n-2}\;dr \nonumber\\
&\approx& 0.059  \;L_{\rm Edd}\times \alpha_{-1}^{0.34}S^{1.2}\lambda^{-1.7}M_{6}^{-0.08}(\beta_{-1}\zeta)^{-0.42}\theta^{0.08}
\int \frac{\Sigma(r,t)}{\Sigma_{0}}r^{n-2}dr 
\label{eq:Lvisc},
\end{eqnarray}
where $L_{\rm Edd}\equiv 4\pi c G m_{\rm p}M/\sigma_{\rm T}$ is the Eddington luminosity
for an object with a mass equal to that of the binary.
The dissipated power 
depends weakly on $\alpha$ and $\theta$,
scales roughly linearly
with the binary mass,
and is most sensitive to the gas distribution parameters $S$ and $\lambda$.
At decoupling, the second integral in equation \eqref{eq:Lvisc} evaluates to a value
$\approx 1/(1-m)$,
and thus the initial luminosity of the disk at decoupling is expected to be
$\sim 0.04 L_{\rm Edd}$ for fiducial parameters and $m\approx -1/2$.
In integrating equation \eqref{eq:Lvisc} before the binary merger,
we do not calculate emission from radii inside $R<2\lambda a(t)$.
In this region, the potential is highly time dependent and non-axisymmetric,
and the assumption that the gas is in nearly circular Keplerian orbits about
the binary's center of mass breaks down.
Hydrodynamical simulations  \citep[e.g.,][]{MM08, Hayasaki+08, Cuadra+09}
suggest that the gas at these radii will form relatively dim, quasi-periodic
accretion flows around the individual MBHs.
We do not address these circumsecondary and circumprimary accretion
flows in this paper.  The potential role
of  circumprimary gas inside of the binary's orbit is discussed by \cite{Chang+09}.

\begin{figure}
\centerline{\hbox{
\plotone{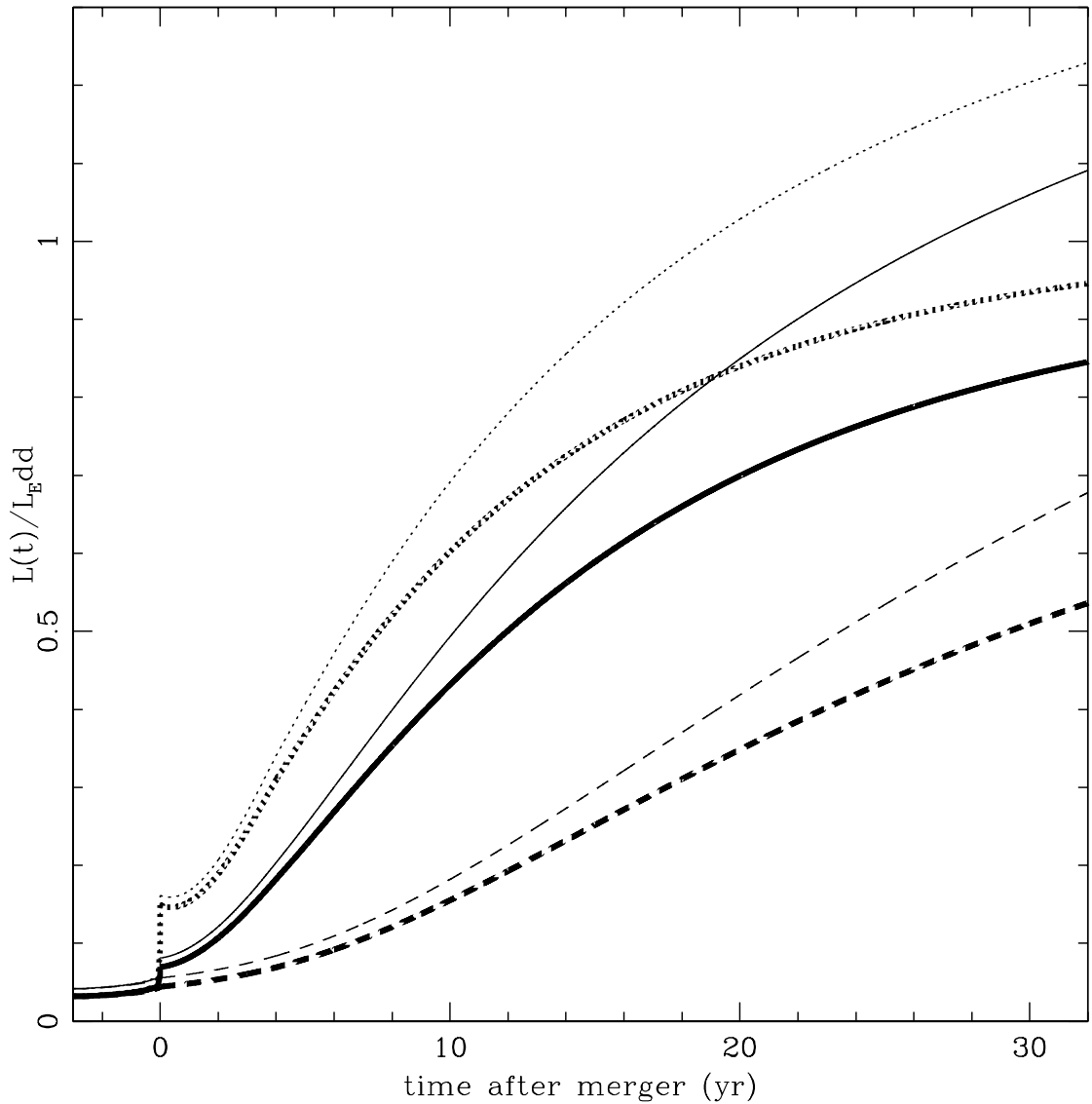}
}}
\caption{
The viscously dissipated bolometric disk luminosity for
fiducial parameters,
in units of the Eddington luminosity.
The time is relative to merger.
The thick lines depict the light curve for density profiles
that have a decretion power-law $m=\dd\ln(\nu\Sigma)/\dd \ln R=-1/2$
outside the inner edge.  The thin lines are for profiles with $m=0$ outside
the inner edge.  The light curves are somewhat sensitive
to the value of $n\equiv\dd\ln\nu/\dd\ln R$, but have the same
qualitative behavior.  The light curves shown here
correspond to disks with viscosity index $n=0.4$
(solid lines), $n=11/170$ (dashed lines) and $n=1$ (dotted lines).
Our model predicts significant evolution before and after
$t_{\rm EM}\sim 9M_{6}^{1.3}\yr$, our re-evaluation of
the nominal time proposed by \citetalias{MP05} for the onset of the
viscosity-powered afterglow.
The sharp increase in luminosity at merger time is a somewhat
artificial effect that arises from the fact that we account only for
the flux dissipated outside the radius $2\lambda a(t)$ (see text),
which shrinks rapidly in the days before merger.
Note that the luminosity scales with the mass-accumulation parameter
$S$ as $L \propto S^{1.2}$,
and so could be significantly enhanced if the inner disk is significantly
more massive than a comparable steady-state Shakura-Sunyaev disk.
}
\label{fig:Lbol}
\end{figure}

In Figure \ref{fig:Lbol} we present several bolometric rest-frame light curves for the evolving disk,
for three different values of the viscosity power-law index $n$,
using equation \eqref{eq:MM08} for the surface density profile at decoupling.
We show light curves for two different surface density profiles
immediately outside the disk edge at decoupling:
a decretion-like disk ($m = -1/2$; thick lines),
and an accretion-like disk ($m=0$; thin lines).
The vertical scale is the brightening $L_{\rm visc}(t)$,
in units of the disk luminosity at decoupling $L_{0}$,
while the horizontal scale is the time relative to the merger.
In each case, the dissipated luminosity increases steadily by
over an order of magnitude within $\sim 20 \yr$ of the merger.
There is significant evolution before and after our re-evaluation of
\citetalias{MP05}'s estimate for the onset for the afterglow,
$t_{\rm EM}=(3/4)t_{\rm GW,0}\sim 9M_{6}^{1.3}\yr$
after the merger.

The sudden brightening that accompanies the merger in
Figure \ref{fig:Lbol} is due to the fact that we only integrate the circumbinary disk
emission outward of the radius $R=2\lambda a(t)$, as described above:
since the binary will merge within $\sim 48$ hours once it reaches
a separation of $\sim 50GM/c^{2}$, the integrated area rises
sharply just before the merger.
This is a somewhat artificial ingredient of our model.
We note, however, that at merger the depth of the potential well near the center of mass
does increase drastically, as does the area where the gas is free to spread viscously.
This sudden change in the central potential well could be associated with
significant emission, for instance if even trace amounts of
gas remain present between the components of the merging binary \citep{Chang+09}.

The disk emission can exceed $L_{\rm Edd}(M)$,
the Eddington luminosity limit associated with the binary mass $M$,
without violating local Eddington flux limits.  This result may appear contradictory, but
it can be understood as follows.
The Eddington flux limit associated with the vertical gravitational field of the disk (ignoring 
self-gravity) is
\beq
F_{{\rm Edd}, z}=\left[1+\left(\frac{H}{R}\right)\right]^{-1/2}\frac{c}{\kappa_{\rm es}}\frac{GM}{R^{2}},
\label{eq:Feddz}
\eeq
where $H$ is the scale height of the disk and
$\kappa_{\rm es}\approx 0.40 \cm^{2}\g^{-1}$ is the electron-scattering
opacity.
The maximum luminosity of a steady disk is then $L_{z}\sim 4\pi \int F_{{\rm Edd}, z}~R~dR$,
where we have introduced an additional factor of two to account for the fact that the disk emits from two faces.
The leading geometric factor in equation \eqref{eq:Feddz} is a monotonically increasing function of $H/R$.
It approaches the value $H/R$ in the limit $H/R\ll 1$, is of order $\sim 0.7$ when $H/R\sim 1$,
and has an asymptotic maximum of unity in the limit $H/R\rightarrow \infty$.
The luminosity $L$ emitted by an annulus with inner and outer radii $R_{\rm in}$
and $R_{\rm out}$ is capped by the inequality
\beq
L<\int_{R_{\rm in}}^{R_{\rm out}}4\pi \frac{cGM}{\kappa_{\rm es}}R^{-1}~dR=\ln\left(\frac{R_{\rm out}}{R_{\rm in}}\right)L_{\rm Edd}.
\eeq
Thus, the bolometric luminosity can reach several Eddington luminosities
for reasonable disk sizes.

We find that the evolution of the light curve is largely determined
by the viscosity profile in the disk
and has a relatively weak dependence on the surface density profile beyond
the disk edge.
Figure \ref{fig:Lbol} suggests that the accretion luminosity of
the viscously spreading inner disk may be a viable
candidate for observational follow-up almost immediately
after the main GW event.
If the electromagnetic accretion signature associated with the spreading inner disk
can be detected in the ``rise'' phase, the light curve
may provide clues about the nature of viscosity in the accretion disk. 
We find that our disk solutions can develop bolometric luminosities
in excess of the Eddington luminosity of the merged MBH remnant.
These luminosities can be generated even if the locally dissipated flux
is sub-Eddington (\S \ref{sec:adv} below), and are enhanced
strongly if the value of the mass-accumulation parameter $S$
greater than unity.

\subsection{Spectral Evolution}
\label{sec:specev}
We now turn to the spectral evolution of the viscously spreading disk.
Below, we present an abbreviated derivation of the thermal structure
of the inner circumbinary disk and discuss its spectral features.
We refer the reader to \citetalias{MP05} and our Appendix \ref{sec:AppA}
for more detailed derivations.

The spectrum differs from that of an exact blackbody because the photons
of different frequencies are thermalized at different depths above the disk.
This thermalizing region is referred to the ``thermalization photosphere''
or the ``effective photosphere.''  The temperature at the bottom of the photosphere
is equal to the effective temperature as seen by an observer far above the photosphere,
viz.  the effective optical depth above this height is of order unity.
We call this temperature the photospheric temperature,  $T_{\rm p}$.

There are two sources of opacity in the photosphere:
electron scattering, which has the frequency-independent opacity 
$\kappa_{\rm es}$;
and absorption, which is dominated by the bound-free process.
We follow \citetalias{MP05} and prescribe a Kramer's functional form
for the absorption opacity
$\kappa_{\rm abs,\nu}\propto \rho T^{-7/2}f_{\nu}(\xi)$,
where $f_{\nu}(\xi)\equiv\xi^{-3}(1-e^{-\xi})$ and $\xi\equiv h\nu/kT$.
We scale $\kappa_{\rm abs,\nu}$ so that its Rosseland mean
recovers the standard Kramer's bound-free opacity for solar
metallicity, $1.6\times 10^{24} (\rho \cm^{3}\g^{-1})(T/\K)^{-7/2}\cm^{2}\g^{-1}$.
For a wide range of system parameters, we find that scattering is the dominant
source of opacity in the photosphere at the inner edge at decoupling.
In this limit, $\kappa_{\rm abs,\nu}\ll \kappa_{\rm es}$, we find
$\kappa_{\rm abs, \nu}\approx\kappa_{\rm abs,*} f_{\nu}^{1/2}$,
with $\kappa_{\rm abs,*}\approx 4.7\times 10^{20 }\cm^{2}\g^{-1}\times 
(\Omega \s) (T_{\rm p}/\K)^{-15/4}$ a frequency-independent quantity.
(See Appendix \ref{sec:AppA} for the  intermediate calculations for
the absorption opacity,
including the general case allowing for $\kappa_{\rm abs,\nu}\sim \kappa_{\rm es}$.)

The so-called ``graybody'' flux emitted by each face of the disk
 is given by the expression \citep[e.g.,][]{RL86, Blaes04}
\beq
F_{\nu}\sim \pi \frac{2\epsilon_{\nu}^{1/2}}{1+\epsilon_{\nu}^{1/2}}B_{\nu},
\label{eq:modF1}
\eeq
where $B_{\nu}(T_{\nu})$ is the Planck function
and $\epsilon_{\nu}\equiv\kappa_{\rm abs,\nu}/(\kappa_{\rm abs,\nu}+\kappa_{\rm es})\le 1$ is the ratio
of the absorption to the total opacity.
Both quantities are evaluated at $T_{\rm p}$, the temperature at the bottom of the photosphere.
Since $\kappa_{\rm abs, \nu}$ is a monotonically decreasing function
of $\nu$, the photosphere emits increasingly less efficiently at higher frequencies,
relative to a blackbody with the same temperature.
In other words, the photosphere has a higher temperature
relative to a blackbody with the same radiant flux, and thus emits
at higher frequencies.
The expression given in \eqref{eq:modF1} recovers the blackbody flux $F_{\nu}=\pi B_{\nu}$ in the limit
where absorption is the dominant source of opacity, i.e. when $\epsilon_{\nu}\rightarrow 1$.
This limit is relevant for our disk solutions, in which $\epsilon_{\nu}$ can span
the full range between 0 and 1 in radii and frequencies of interest.
Note that the flux expression used by \citetalias{MP05} is lower than ours by a factor of two.

If again advective losses are ignored, integrating the flux
of equation \eqref{eq:modF1} over frequency must give the
flux viscously dissipated by half of the disk, $(9/8)\nu\Sigma \Omega^{2}$.  We obtain
\beq
\frac{F_{\rm visc}}{2}
=\int_{0}^{\infty}F_{\nu}\;d\nu
=\Xi(T_{\rm p},\Omega)\, \sigma T_{\rm p}^{4}
=\frac{9}{8}\nu\Sigma\Omega^{2},
\label{eq:fluxeq1}
\eeq
where
\beq
\Xi(T_{\rm p},\Omega)\equiv
\frac{15}{\pi^{4}}
\int_{0}^{\infty}\frac{2\epsilon_{\nu}^{1/2}}{1+\epsilon_{\nu}^{1/2}(\xi)}
\frac{e^{-\xi}d\xi}{f_{\nu}(\xi)}
\approx
\frac{4}{5}\epsilon_{*},
\eeq
is the deviation of the bolometric flux from blackbody and 
$\epsilon_{*}(T_{\rm p},\Omega) \equiv \kappa_{\rm abs,*}/(\kappa_{\rm abs,*}+\kappa_{\rm es})< 1$.
The above approximation for $\Xi$ is accurate to within $10\%$ in the range $0\le \epsilon_{*}\ltsim 0.9$.
(See Appendix \ref{sec:AppA} for a more accurate fit to the integral.)

To calculate the spectrum, we solve equation \eqref{eq:fluxeq1}
numerically for $T_{\rm p}$, substitute the result into 
equation \eqref{eq:modF1} and integrate:
\beq
L_{\nu}(R,t)\approx 2\pi^{2} R_{0}^{2}\int
\frac{2\epsilon_{\nu}^{1/2}(\Omega, T_{\rm p})}{1+\epsilon_{\nu}^{1/2}(\Omega, T_{\rm p})}
B_{\nu}(T_{\rm p}) \;r\;dr.
\label{eq:spec}
\eeq

The temperatures at the midplane and the bottom of the photosphere
are related through
\beq
T^{4}=\frac{3\tau}{4}T_{\rm p}^{4},
\label{eq:TTp}
\eeq
if the region between the two heights is optically thick and
can be treated as a one-zone gray atmosphere.
Here $\tau=\theta\kappa_{\rm es}\Sigma$
is the scattering-dominated optical thickness between the two heights,
and $\theta\le 1$ is a porosity correction factor.
We adopt \citetalias{MP05}'s interpretation of the simulation results
of \cite{Turner04} and use $\theta=0.2$ as our fiducial value,
and define $\theta_{0.2}\equiv\theta/0.2$.

Neglecting the advected flux, and in the limit where the circumbinary
disk is dominated by radiation pressure
and electron scattering, we may estimate the
frequency for the peak monochromatic flux at decoupling,
and at late times after the merger.
At decoupling, most of the emission comes from the inner edge,
and the spectrum in equation \eqref{eq:spec} peaks at the Wien frequency
$\xi=h\nu/kT_{\rm p,0}\approx2.8$, from which we estimate
\beq
h\nu_{\rm peak}\approx 15 \eV \times
\alpha_{-1}^{0.36}S^{0.73}\lambda^{2.1}M_{6}^{-0.32}(\beta_{-1}\zeta)^{-0.45}\theta_{0.2}^{0.09}.
\label{eq:nupeak}
\eeq
Long after the merger, the inner disk approaches a quasi-steady accretion
track satisfying $3\pi\nu\Sigma=\dot{M_{\rm out}}(1-\sqrt{R_{*}/R})$,
where $\dot{M_{\rm out}}$ is a radially constant mass supply rate of the inner disk.
The bolometric emission is brightest where the quantity
$2\pi R\nu\Sigma \Omega^{2}$ is maximal in the disk,
which corresponds to $R=25/16 R_{*}$.  Applying this to equation \eqref{eq:fluxeq1},
we calculate the photospheric temperature and find the corresponding peak frequency
for the monochromatic flux at very late times to be
\beq
h\nu_{\rm peak}\approx 0.71 \keV \times M_{6}^{-8/34}\left(\frac{\dot{M}}{\dot{M}_{\rm Edd}}\frac{0.1}{\eta}\right)^{8/17}
\left(\frac{R_{*}c^{2}}{3GM}\right)^{-18/17}.
\eeq
Above, the dependence on $S$ is replaced by a dependence on the mass
supply rate, as at late times the disk loses memory of earlier
accumulation near the inner edge.
For our choice of $R_{*}= 3GM/c^{2}$,
the radiative accretion efficiency is
$\eta=1-\sqrt{1-2GM/(R_{*}c^{2})}\approx 0.12$ \citep{NT73}.
The circumbinary disk is expected to reach this quasi-steady track
after at least $t_{\rm visc,0}\approx 120 M_{6}^{1.3} \yr$ has elapsed since merger.

Figure \ref{fig:sp} shows the spectrum of our fiducial model disk around an equal-mass,
$10^{6}\Msol$ binary as calculated at different times with respect to the merger, from
decoupling time ($\approx 3 \yr$ before the merger)
up to $120 \yr$ after the merger.
We show the exact results using equation \eqref{eq:spec}
in solid lines\footnote{
The function $\nu F_{\nu}$ peaks at $\xi=3.9$, so replace the leading factor $15 \eV$ in
equation \eqref{eq:nupeak} above with $21 \eV$ to derive the locations of the peaks in the plots of Figure \ref{fig:sp}.
}
, and the blackbody spectrum in dotted lines.
In the figure, we only account
for emission out to a radius of $1000 GM/c^{2}$,
as we do not expect significant contribution of high-frequency photons
from regions far outside $R_{0}$.
We find that the spectrum of the circumbinary disk does not evolve significantly from that
at decoupling until several hours before the merger, because
the deepest parts of the central potential are not accessible to the circumbinary
gas until that time.
Our initial and final spectra agree qualitatively with those given in \citetalias{MP05}.
As with the bolometric light curve, however, significant evolution is ongoing
well before and after the nominal estimate for the onset of the afterglow.
Much of the evolution, however, is complete $\sim 10M_{6}^{1.3}\yr$ after the merger.

\subsection{Possible Reprocessing of the X-ray Signature}
The frequencies calculated above suggest that an evolving
afterglow at low redshift or low binary mass
may be observable by existing and future
X-ray and UV telescopes shortly after the GW event.
\citetalias{MP05} discussed the possibility that the afterglow
could be reprocessed to IR frequencies if the
merged binary is enshrouded by gas and dust.
Here we note that the circumbinary disk itself is
a plausible candidate to reprocess the UV/X-ray afterglow to
lower frequencies.

Ignoring general relativistic effects, the self-irradiating flux of
the disk with viscously dissipated flux $F_{\rm visc}$ can be written
\citep[see, e.g., ][]{Blaes04}
\beq
F_{\rm irr}(R)=\frac{1-A}{4\pi R^{2}}\int 2\pi R^{\prime}\sin\vartheta F_{\rm visc}(R^{\prime}) \; dR^{\prime},
\label{eq:repro}
\eeq
where $A$ is the albedo of the disk and $\vartheta$ is the angle
between the disk surface and the irradiating light ray.
In our disk solutions, most of the dissipated disk luminosity $L_{\rm visc}$ is produced at small radii due
to the fact that the viscously dissipated flux is proportional to $\Omega^{2}$.
Thus, if $R$ is much greater than the size of the central emitting region then
it is a reasonable approximation to replace the integral in the above equation
with $L_{\rm visc}\sin\vartheta$.  If one further assumes that $A$ is small
and that the irradiating flux is large compared to the local value of $F_{\rm visc}$,
the luminosity reprocessed by the outer disk
will be $L_{\rm irr }\ltsim L_{\rm visc}\sin\vartheta.$
The light-travel time is short ($\sim 10^{-2} M_{6} \yr$ across distances $\sim 10^{5}GM/c^{2}$)
compared to the timescale for viscous evolution of the
inner disk, and thus we expect the reprocessing processes below
to be effectively instantaneous.

One way that the factor $\sin \vartheta$ can be large is if there is a warp
in the disk that offsets the planes of the outer and the innermost regions.
The circumbinary disk is expected to be aligned with the binary orbit
in the binary's vicinity, where tidal torques dominate viscous torques.
Farther out in the disk, the disk may have a quasi-steady warp
with respect to the binary \citep{IPP99}.
Because the warp would dissipate on timescales longer than
the afterglow itself, it could act to promptly reprocess the afterglow into longer wavelengths.

We also find that the evolution of the innermost accretion flow itself
could effect a disk geometry that is conducive to reprocessing.
If there is no warp and the emitting region is in the equatorial plane,
then \citep[e.g.][]{Blaes04}
\beq
\sin\vartheta \sim \frac{H}{R}\left(\frac{d \ln H}{d\ln R}-1\right).
\eeq
As we show below (Figures \ref{fig:adv1} and \ref{fig:adv2}), just outside
the regions of brightest emission ($R\ltsim R_{0}$)
and around the time of merger 
our disk solutions are often moderately geometrically thick ($H/R\gtrsim 0.1$)
and steeply flared geometrically (large positive $d\ln H/d\ln R$).
That is, the circumbinary disk itself may serve as a
shroud of gas that can capture and reprocess the high-frequency accretion signature.

Either of the two effects considered above could
cause the luminous inner disk around the binary remnant
to irradiate the outer disk at a sufficiently large angle $\vartheta$
to effect a reprocessed IR/optical signature of substantial
luminosity.

\subsection{Possible Effects of Advection and Super-Eddington Winds}
\label{sec:adv}
Above, we largely confirmed the findings of \citetalias{MP05},
with the additional suggestion that accretion afterglows
of MBH mergers may be observable somewhat earlier than estimated in that study.
However, we also find that at late times the spreading accretion flow
becomes geometrically thick near the MBH remnant,
with the scale-height-to-radius 
ratio $H/R$ formally exceeding unity.  Here, $H=\Sigma/\rho=c_{\rm s}/\Omega$,
where $c_{\rm s}$ is the sound speed $\sqrt{\gamma P/\rho}$
and $\gamma\approx 4/3$ is the adiabatic index for a radiation-pressure-dominated gas.
The radial profile of $\rho$ is calculated from the disk temperature profile,
which is obtained through equations \eqref{eq:fluxeq1} and \eqref{eq:TTp}
and the total pressure in the disk $P_{\rm gas}(\rho,T)+P_{\rm rad}(T)$.
 
For some binary and disk parameters,
the inner edge of the disk may already be geometrically thick
at decoupling (\citetalias{MP05}; see Table \ref{tab:1}).
In such a scenario, the
circumbinary gas may be free to accrete off-binary plane,
and the inner cavity may not be as evacuated as we have assumed here,
and the difference between the pre- and post-merger disk spectra
may be less easily discernible.
In addition, after the merger, it is also possible for the accreting gas to become radiatively inefficient,
as in the advection-dominated accretion flow \citep[ADAF; ][]{NY94, Abramowicz+88} model,
which would make it more difficult to observe\footnote
{Here, we mean a single-temperature, collisional, radiatively inefficient flow
(as opposed to the collisionless, optically thin, two-temperature models in the literature, e.g.,
\citealt{Shapiro+76, Rees+82, NY95b}).
}.

As argued in \citetalias{MP05}, if horizontal advection is competitive with
viscous dissipation then the physical disk would be thinner than
suggested by our simple estimate of the ratio $H/R$,
as in the ``slim disk'' model of \cite{Abramowicz+88}.
In this paper, we do not attempt to develop a self-consistent 
time-dependent model that incorporates vertical disk structure and radial advection.
Rather, we proceed below as if the disk evolution can be approximated
by the thin-disk evolution described by equation \eqref{eq:Green} while keeping
the above caveats in mind.

The absolute value of the advected flux can be estimated as \citep[e.g.,][]{Abramowicz+88}
\beq
\mathcal{Q}_{\rm adv}=\frac{\dot{M}}{\pi R^{2}}\frac{P_{\rm rad}}{\rho}
\left|\frac{d \ln \rho}{d \ln R}-3\frac{d\ln T}{d \ln R}\right|.
\label{eq:Qadv}
\eeq
Substituting equation \eqref{eq:Mdot} for the local accretion rate and
equation \eqref{eq:fluxeq} for the radiated flux, we crudely estimate
the advected to radiated flux ratio as:
\beq
\frac{\mathcal{Q}_{\rm adv}}{\mathcal{Q}_{\rm rad}}\approx\frac{4}{3}\frac{P_{\rm rad,}H}{\Sigma R^{2}\Omega^{2}}
\left[1+\frac{d\ln \left(\nu\Sigma\right)}{d\ln R}\right]\left|\frac{d \ln \rho}{d \ln R}-3\frac{d\ln T}{d \ln R}\right|
\label{eq:Qrat}
\eeq

\begin{figure}
\centerline{\hbox{
\plotone{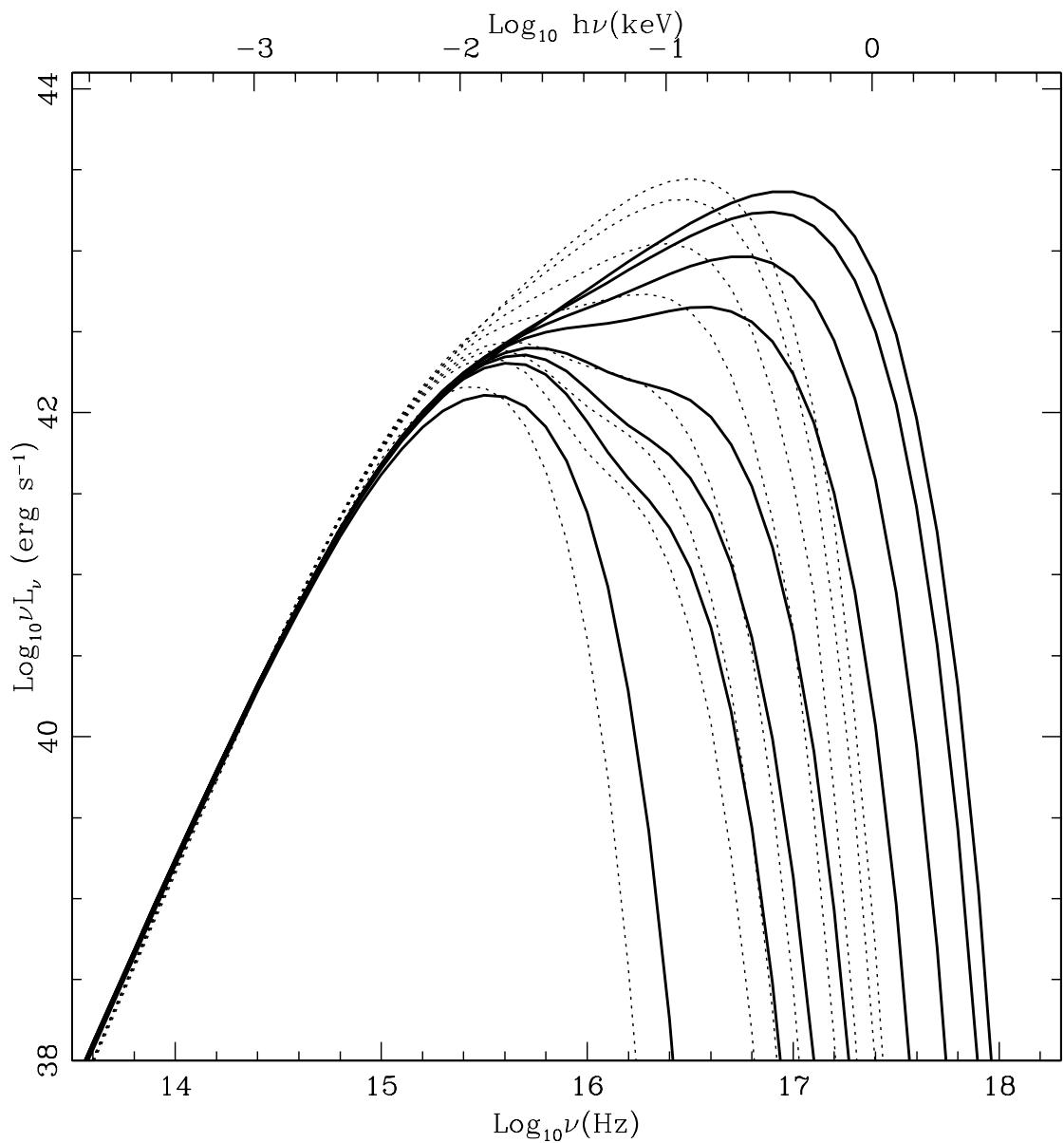}
}}
\caption{
The thermal disk spectra emitted at (from left to right):
decoupling (3.1 years before the merger), 1 month after,
1 year after, 2 years after,
5 years after, 9.2 years after, 20 years after, and 120 years after the merger.
The thick, solid lines show the
spectra calculated using the ``graybody'' formulation in
equation \eqref{eq:spec}, while the thin, dotted lines show
the spectra calculated assuming the disk instead emits as a perfect blackbody.
 For any given snapshot, the two sets of lines show different spectral distributions
of the same bolometric luminosity.
For the gray-body case, 
the 9.2-year snapshot corresponds to our re-evaluation of the nominal estimate by \citetalias{MP05}
for the onset of the X-ray afterglow, and 120 years is the
viscous time at the disk edge at decoupling.
}
\label{fig:sp}
\end{figure}

The simple calculations of the evolving disk spectrum performed in 
\S \ref{sec:specev} are not highly dependent on the disk temperature
(our Figure \ref{fig:sp} agrees
very well with the spectra shown in \citetalias{MP05},
despite the fact that we have calculated a somewhat lower
value of $T_{0}$).  However, the structure of the disk
at the inner edge at decoupling and the subsequent
evolution of the accretion flow is
likely to depend much more sensitively on the temperature profile.
In the radiation-dominated limit, $H$ is proportional to $P_{\rm rad}\Sigma^{-1}\Omega^{-2}$
so from equation \eqref{eq:Qadv} we see that
the advected flux is roughly proportional to
$\nu T^{8} \Sigma^{-1}\Omega^{-2} R^{-2}$.
This implies that near the inner edge, $\mathcal{Q}_{\rm adv}\propto T^{62/7}$ approximately (see Appendix
\ref{sec:AppA} for how disk quantities such as $\Omega$ scale with the temperature near the edge).
Thus, we expect the possible transition of the 
disk from a thin disk to an advective/slim disk or a geometrically thick ADAF
$-$ and the corresponding spectral evolution $-$ to be
a delicate function of the thermal structure of the disk at decoupling.
For example, \citetalias{MP05} estimated a midplane temperature
at the inner edge that was $\sim 1.3$ times higher than we did,
with most of the discrepancy due to the aforementioned factor of two in the
flux formalization.  Although this does not result in significant
differences between our values and theirs for
the decoupling radius or the relevant timescales,
it does lead those authors to estimate that the advective
flux is more significant at the disk edge at decoupling
than we do.  We estimate that immediately outside the disk edge,
\beq
\frac{\mathcal{Q}_{\rm adv}}{\mathcal{Q}_{\rm rad}}\gtrsim 10^{-2}
\alpha_{-1}^{1.5}S^{4.9}\lambda^{-7.6}M_{6}^{-0.24}\left(\beta_{-1}\zeta\right)^{-1.9}\theta_{0.2}^{2.4},
\eeq
where \citetalias{MP05} estimated $\mathcal{Q}_{\rm adv}/\mathcal{Q}_{\rm rad}\sim 0.44$.
The flux ratio is higher at the edge due to the steep temperature
and density gradients there.  The equation above suggests that it is highly sensitive to all of the disk
and binary parameters save for the binary mass.

\begin{figure}
\centerline{\hbox{
\plotone{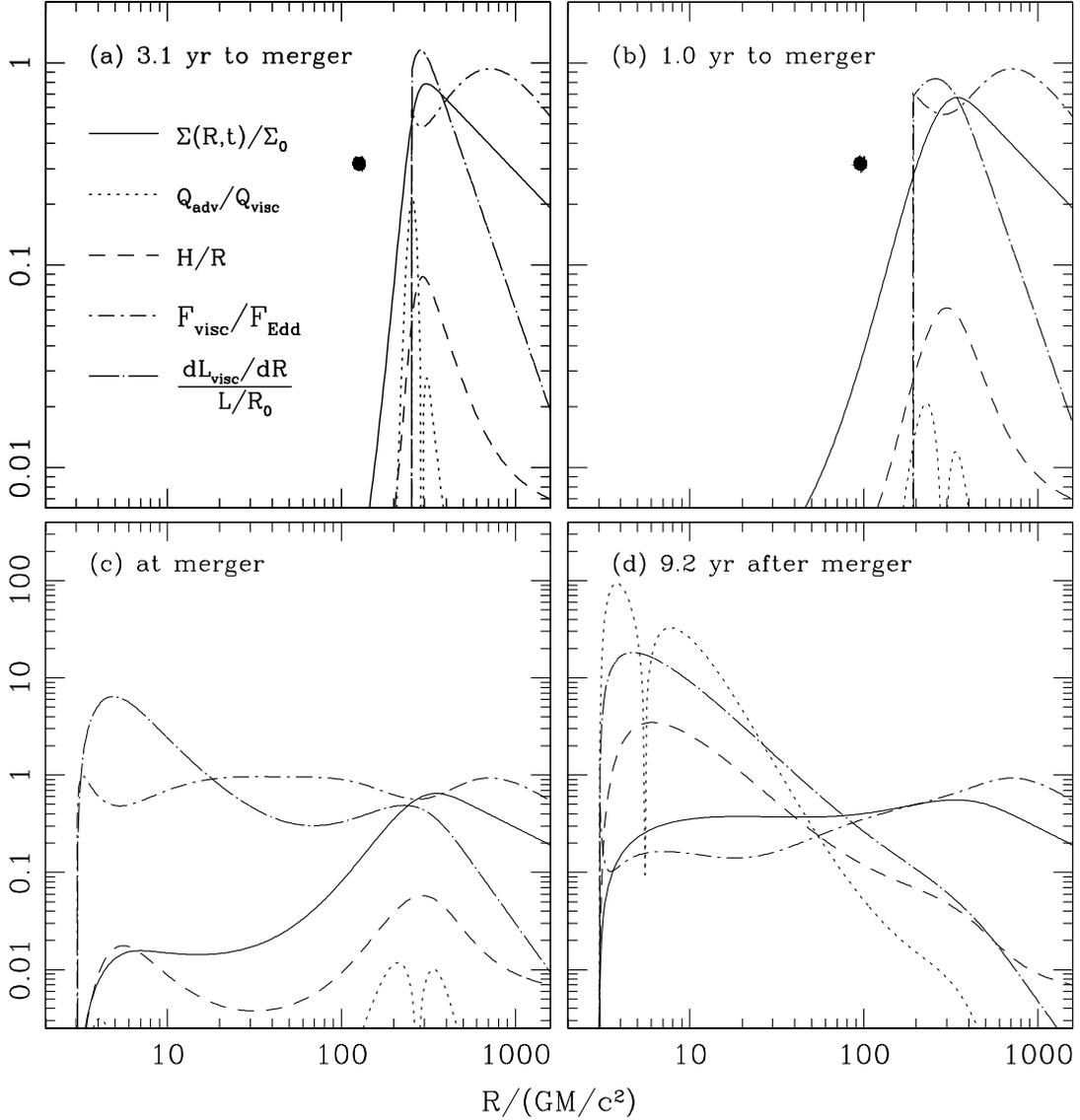}
}}
\caption{
For our fiducial disk model:
the surface density, scale-height-to-radius ratio $H/R$;
estimated ratio of the advected to radiated flux $\mathcal{Q}_{\rm adv}/\mathcal{Q}_{\rm rad}$;
the ratio of the local flux to the Eddington limit $F_{\rm visc}/F_{\rm Edd}$;
and the flux contribution $dL_{\rm visc}/dR=2\pi R F_{\rm visc}$ normalized to $L/R_{0}$.
The disk is geometrically thin and advection is largely insignificant until the
merger.  After the merger, however, the inner disk becomes formally geometrically thick
and advection-dominated (Panel d).
Note that in panels (c) and (d) the vertical scale is different from the other two panels.
The steep dips in $\mathcal{Q}_{\rm adv}/\mathcal{Q}_{\rm rad}$ are due to
the advected flux changing sign (see equation [\ref{eq:Qrat}]).
}
\label{fig:adv1}
\end{figure}

In Figure \ref{fig:adv1} we show the radial profiles of 
several additional disk quantities for the same parameters, profile and times
as shown in Figure \ref{fig:sig1}.
Along with the surface density, we also plot:
 the simple estimate from equation \eqref{eq:Qrat} for the advected-to-radiated flux ratio,
$\mathcal{Q}_{\rm adv}/\mathcal{Q}_{\rm rad}$;
the estimated scale-height-to-radius ratio $H/R$; 
and $F/F_{\rm Edd}$, the ratio of the locally dissipated
flux to $F_{\rm Edd}\approx H\Omega^{2}c/\kappa_{\rm es}$,
the local Eddington flux limit
associated with the vertical component of gravity.
For the last ratio, we find
\beq
\frac{F}{F_{\rm Edd}}
\approx 0.06 \frac{\Sigma(R,t)}{\Sigma_{0}}\sqrt{1+(H/R)^{-2}}r^{n-1}
\alpha_{-1}^{0.34}S^{1.2}\lambda^{-1.70}
M_{6}^{-0.08}\left(\beta_{-1}\zeta\right)^{-0.42}\theta_{0.2}^{0.08}.
\eeq
Figure \ref{fig:adv1} confirms that for fiducial parameters,
the circumbinary accretion flow remains locally sub-Eddington.
For fiducial parameter values, we find that a relatively high surface density
$\Sigma/\Sigma_{0}\gtrsim 1$ (or large $S$)
is required to reach a super-Eddington flux near $R=R_{*}$, and
the disk is thus not necessarily likely to develop super-Eddington winds at levels
which would significantly affect the light curves or spectra presented in Figures \ref{fig:Lbol} and \ref{fig:sp}.
At late times, however, the nominal advective flux
exceeds the locally dissipated flux as the disk becomes geometrically thick,
which may lead the disk to become radiatively inefficient.

\begin{figure}
\centerline{\hbox{
\plotone{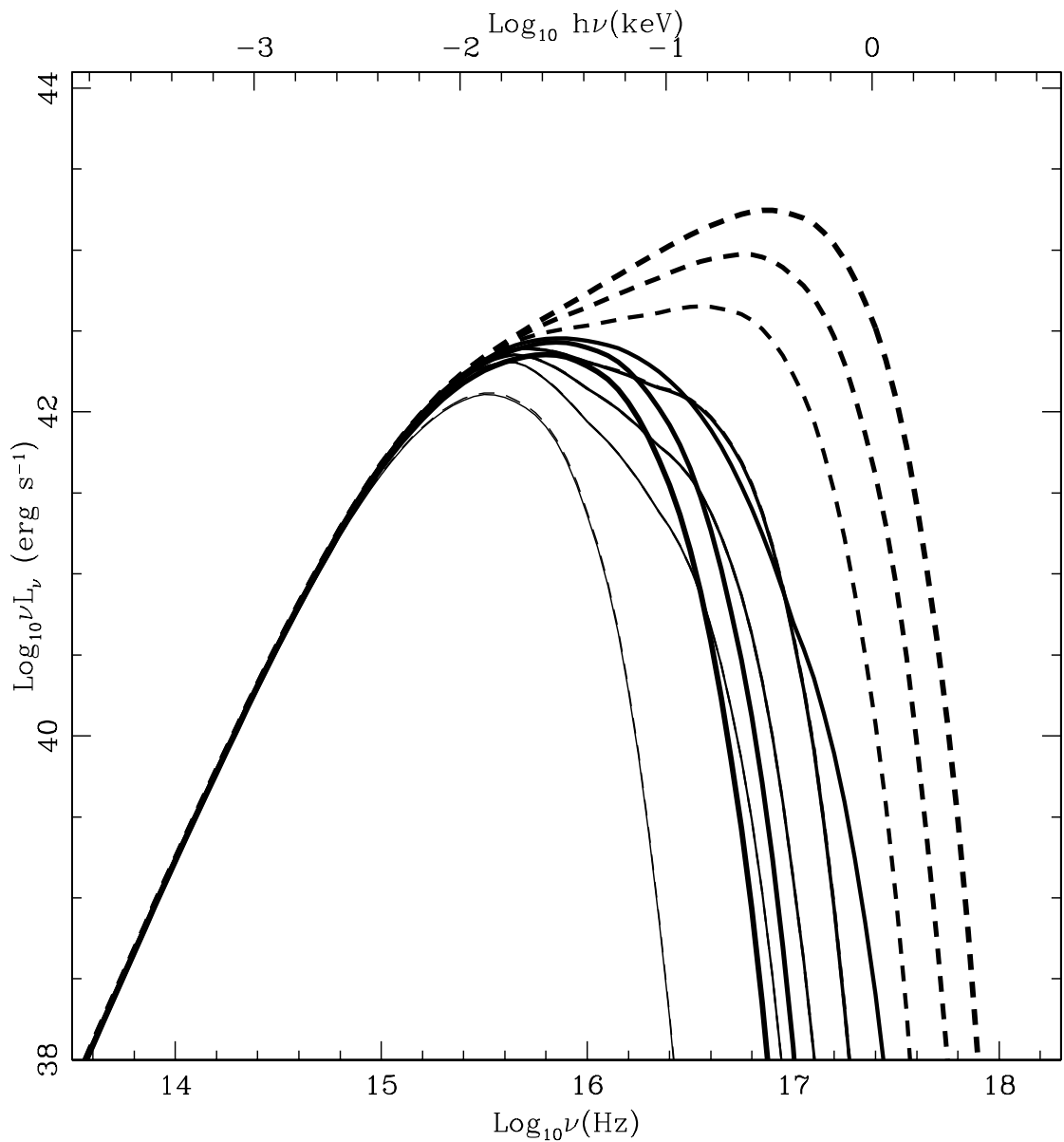}
}}
\caption{
Emission spectra for the fiducial circumbinary disk.
The dashed lines show the same spectra as calculated in Figure \ref{fig:sp},
from thinnest to thickest curves,
at decoupling and at 1 month, 2 yr, 5 yr, 9.2 yr and 20 yr after the merger.
The solid lines show the spectra at the same times, but after subtracting
the estimated advected flux.
The disk is initially geometrically thin and advection is not significant
until after the merger.  At late times, the inner disk becomes geometrically
thick and radiatively inefficient, and this could lead to reduced overall X-ray emission.
The disk emits briefly emits soft X-rays at intermediate times, $\sim 2\yr$
after the merger.
}
\label{fig:advsp1}
\end{figure}

This hypothetical transition of the innermost accretion flow from 
a thin disk to a geometrically thick flow could
act to suppress the X-ray afterglow.
We estimate very simply the effect of advection on the emitted spectrum as follows.
We assume that the emitted flux is
suppressed by a factor $f\equiv \max\{0, 1-\mathcal{Q}_{\rm adv}/\mathcal{Q}_{\rm rad}\}\le 1$,
and that the advected energy is not re-emitted.
We also assume that advection acts to suppress emission
evenly across all photon frequencies.
The resulting advection-limited spectrum is shown in Figure \ref{fig:advsp1}
alongside a spectrum calculated without accounting for advection
(i.e. the same disk spectra as those shown in
Figure \ref{fig:sp}).
from decoupling to 20 years after the merger.
We compare the two classes of spectra
at decoupling, and 1 month, 2 yr, 5 yr, 9.2 yr and 20 yr after the merger,
with thicker lines in the figure denoting later times.
If the disk were to remain radiatively efficient at all radii and times,
the emission at high frequencies would steadily increase until the quasi-steady
thin-disk track is reached.  If instead the innermost disk becomes
advective, this suppresses the high-energy emission at late times.
Our results indicate
that our fiducial disk model may become increasingly less bright
as the inner disk becomes increasingly geometrically thick.
However, the calculations shown in Figure \ref{fig:advsp1}
suggest that a brightening afterglow may still be visible in the soft X-ray
for a short time $\sim 10^{-2}t_{\rm visc,0}$ after
the merger.
This suggests that the circumbinary gas
may have two observable signatures of interest:
an initial brightening phase with a soft X-ray afterglow,
and a subsequent dimming phase as it transitions into an ADAF.

To further investigate the possible consequences of advection on the circumbinary disk evolution,
we turn to a different disk model whose inner edge is already geometrically thick
at decoupling.
We take $S=5$ and $\zeta=1/3$ (binary mass ratio $q\approx 1/10$), keeping all other parameters
the same as in the fiducial model.  The new choices for $S$ and $\zeta$
are physically plausible ones.
If the mass supply rate of the outer
disk is comparable to the Eddington limit of the central
binary, we can expect $S>1$ at decoupling as a result of mass accumulation.
We may also expect
$q\sim 1/10$ mergers to be more common\footnote{
Provided that they can overcome the ``last-parsec problem.''}
than near-equal-mass mergers because merging galaxies (dark matter halos)
have unequal masses and because the mass contrast between the MBHs is expected
to be higher than that of the host halos
(i.e. the observationally inferred MBH-to-host-halo
mass ratio relation is a steeper-than-linear
function of the halo mass; \citealt{Ferrarese02}).
Cosmological merger tree calculations for MBH mergers consistently show that contributions to the
{\it LISA} data stream from MBH coalescences
will be dominated by moderately low-$q$ events
\citep[e.g.,][]{Volonteri+03, Sesana+07, TH09}.

\begin{figure}
\centerline{\hbox{
\plotone{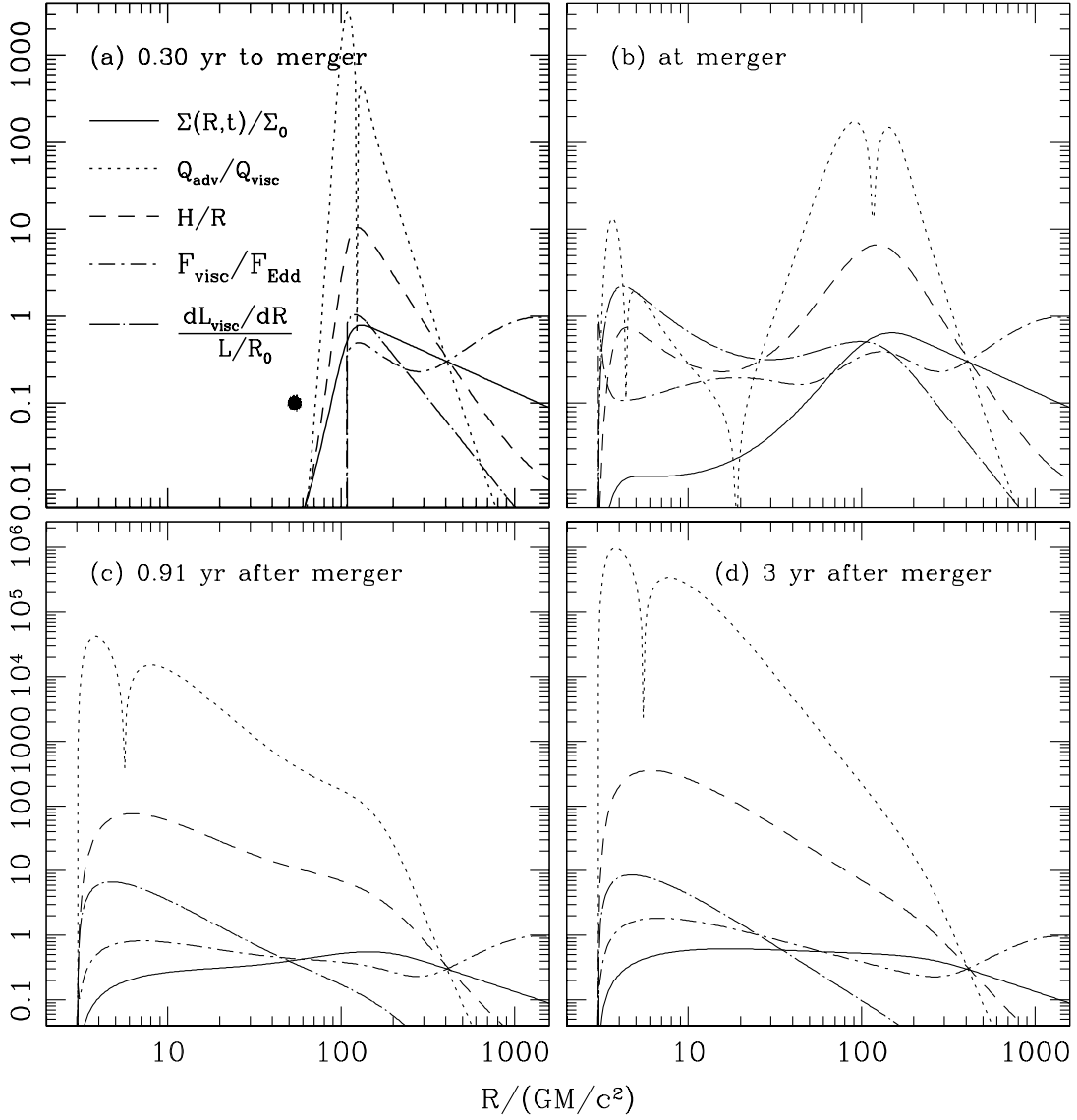}
}}
\caption{
Same as Figure \ref{fig:adv1}, but for a more massive and
geometrically thicker disk with $S=5$ and $\zeta=1/3$.
The disk is geometrically thick and advection-dominated in the inner regions
of interest, from decoupling to several years after the merger.
The local flux also exceeds the Eddington limit at late times (Panel d).
}
\label{fig:adv2}
\end{figure}

\begin{figure}
\centerline{\hbox{
\plotone{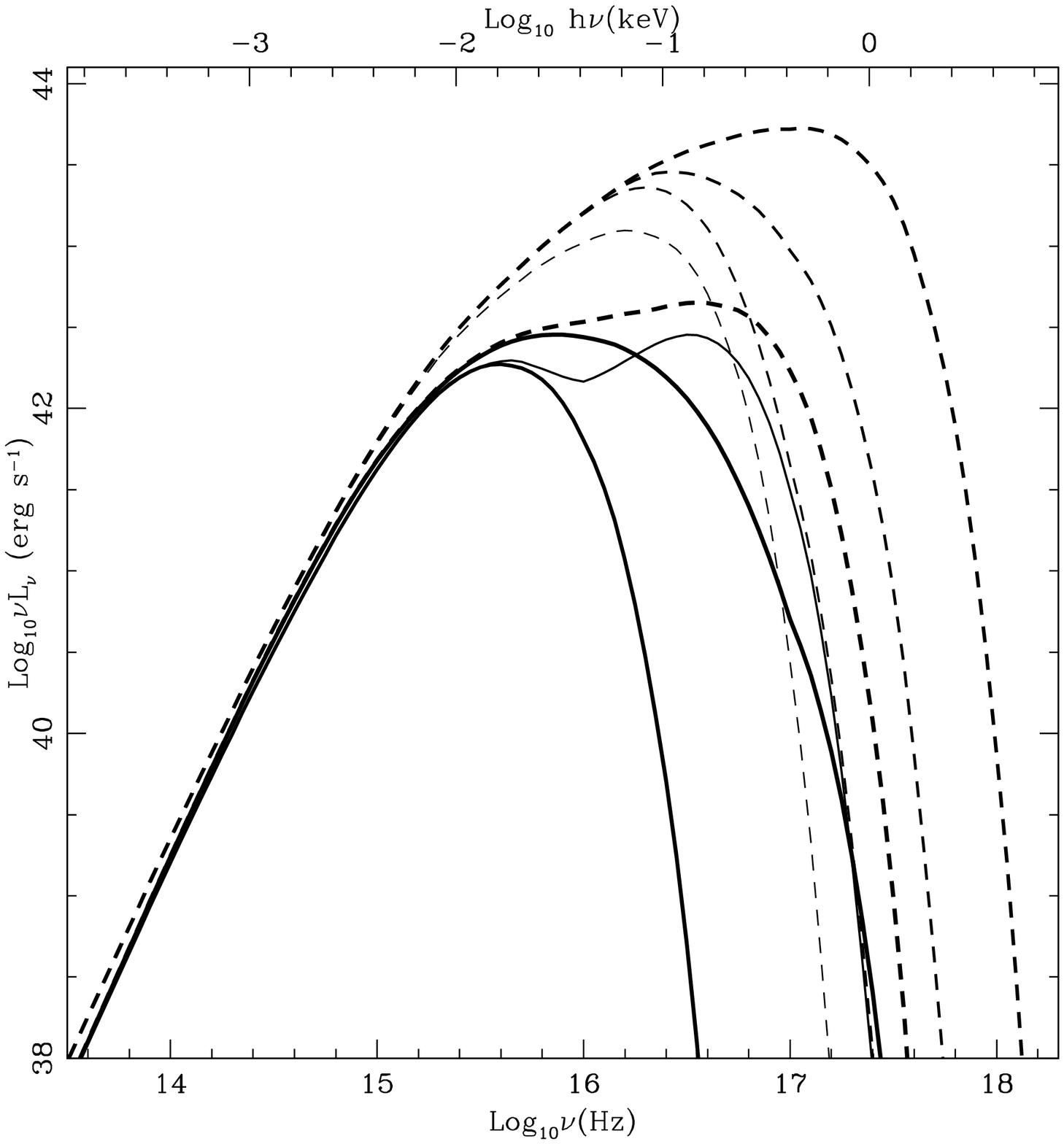}
}}
\caption{
The spectra of a more massive and geometrically thick disk,
with $S=5$ and $\zeta=1/3$, at decoupling
and at 1 week, 3 months, $0.91 \yr=t_{\rm EM}$
and $3 \yr$ after the merger.
As in Figure \ref{fig:advsp1}, later spectra
are shown with thicker lines.  The dashed lines show
spectra ignoring advective effects, while the solid lines
show spectra under the assumption that advection suppresses the emitted flux.
In the latter case, the disk is geometrically thick and radiatively inefficient
across the entire region inside the initial gap radius,
 and thus no obvious afterglow may be observable (all curves overlap except
 the one 1 week after merger, when our calculations show a brief afterglow that
 is quickly suppressed by advection).
}
\label{fig:advsp2}
\end{figure}

With this new set of parameters, the disk decouples at a time
$t_{\rm GW,0}/4\approx 0.30 \yr$ before the merger,
when the binary has reached a semimajor axis
$a_{0}\approx 54 GM/c^{2}$.
The nominal estimate for the onset of the afterglow emission
is $t_{\rm EM}\approx 0.91\yr$ after the merger.
The inner edge of the disk has a somewhat higher
temperature, $T_{0}\approx 8.7\times 10^{6}\K$.
We again use the initial density profile given by equation \eqref{eq:MM08}.
In Figure \ref{fig:adv2} we show the same disk quantities
for this thick disk as shown for the fiducial disk in Figure \ref{fig:adv1}.
This time, the disk remains geometrically thick and radiatively
inefficient inside the initial edge radius throughout its evolution.
As we did for the fiducial disk model, we estimate
the advective suppression of the spectrum for the thick disk
and show results in Figure \ref{fig:advsp2}.
This simple calculation suggests that a disk that
is sufficiently geometrically thick at decoupling
may not, in principle, exhibit any observable spectral evolution.

On the other hand, we also find that the viscously dissipated flux
of this thick circumbinary disk becomes super-Eddington at late times
(Figure \ref{fig:advsp2}d).
\cite{Begelman02} proposed that accretion disks with a super-Eddington flux
may be able to stay geometrically thin and radiatively efficient.
Such accretion flows may also generate powerful outflows and flares.
While our simple models do not self-consistently
treat advection or the super-Eddington regime,
our results underscore the need for further study of this class
of accreting systems.

\section{Conclusion}
\label{sec:concl}
We have presented a simple semianalytic model
for the viscous evolution of a thin circumbinary disk
around a MBH binary, in the final stages of the binary's evolution.
Using this model, we have estimated the time dependence
of the thermal spectrum immediately after the binary merger.
In what may be the most optimistic scenario,
a rapidly evolving soft X-ray signature
may be observed soon after the GW event, perhaps
years earlier than previously estimated.
It is worth emphasizing that 
the bolometric luminosity is several orders of magnitude brighter
than many of the EM counterpart candidates proposed in the literature,
and in extreme cases may exceed the Eddington luminosity of the central MBH.
An important feature of the afterglow mechanism discussed in this paper
is that it is not strongly limited by the mass of the circumbinary disk.
This is in stark contrast to recoil-powered afterglows
whose luminosities and observational prospects are generally limited by the mass of the gas
bound to the recoiling remnant
\citep[e.g.,][]{BP07, SK08, Lippai+08, SB08, O'Neill+09, Megevand+09, Corrales+09, Rossi+09, Anderson+09}.

The reason why an afterglow may be observable so early is because the
power per unit area dissipated by viscous dissipation scales as $F\propto \nu\Sigma\Omega^{2}$.
The potential is sufficiently deep close to the MBH remnant that
in this region even low surface densities can generate more power than
elsewhere in the disk.
A central conclusion of this paper is that
enough gas may be able to viscously follow the
binary to merger.
The presence of gas in the vicinity of the binary
at merger may have significant implications for other proposed
EM counterpart mechanisms, as well.  Some of this gas can reasonably
be expected to contribute to circumprimary or circumsecondary
accretion disks which could generate additional observable signatures even if
the total gas mass captured in this way is small \citep{AN05, Lodato+09, Chang+09}.

In addition to the possibility that the accretion afterglow proposed
by \citetalias{MP05} may be observable earlier than previously estimated,
our results raise interesting possibilities about the observable features
of the post-merger disk at later times. 
Most of the luminosity is generated in the innermost disk ($R\ltsim 20 GM/c^{2}$), and
may be partially reprocessed to IR/optical frequencies
by the outer disk ($R\gtrsim 100 GM/c^{2}$).  We have considered two mechanisms
that could reasonably reprocess a significant fraction of the inner-disk emission:
a geometrical warp in the disk just outside the emitting region; and
the steep geometrical flaring of the vertical thickness of the inner disk as it evolves
We find that the bolometric luminosity
of the time-dependent afterglow can exceed the Eddington luminosity
of the binary, without violating the local Eddington flux limit.
We conclude that the EM signature --- the unobstructed signature in
soft X-rays, as well as the reprocessed signature in IR/optical ---
could become comparable
to the Eddington luminosity of the binary.

Our calculations also suggest that as it spreads,
the innermost disk may become
geometrically thick and thus radiatively inefficient,
even if it is neither geometrically thick nor radiatively inefficient at decoupling.
If the disk is initially geometrically thin at decoupling, then
it may emit an afterglow that is powered by the newly formed deep
central potential before becoming geometrically thick
near the remnant at later times.  In such a scenario, an
accretion disk around a recently merged MBH remnant
may provide a unique system where the transition from
a radiatively efficient to an inefficient accretion state can
be monitored on humanly tractable timescales.

In the most pessimistic scenario, it seems possible for the disk
to become geometrically thick even before the binary merges.
In such a situation a disk could behave like an ADAF and it is unclear whether a circumbinary cavity
should exist at all.  Even if a cavity were kept open until decoupling, advective losses
may suppress any observable spectral evolution for the disk emission.
As suggested by \citetalias{MP05},
horizontal advection could also act to
make the disk thinner than we have considered here, as in the
``slim disk'' models of \cite{Abramowicz+88}.  A slim disk
would remain somewhat radiatively efficient and thus could still
exhibit an observable evolution of the spectrum or luminosity.
Another possible mechanism to keep the circumbinary disk radiatively
efficient is super-Eddington fluxes \citep{Begelman02}.
We find that in the same regions of the parameter space
where the disk becomes very geometrically thick,
the viscously dissipated flux also becomes very high.
The nominally radiated fluxes near the central regions around the MBH remnant
can be super-Eddington for physically reasonable parameter values,
and this could also help the disk produce an observable
evolving EM signature.

Advection-dominated accretion or a super-Eddington flow
may produce powerful outflows
\citep[e.g.,][]{BB99}.  The outright viscous dissipation of super-Eddington fluxes
near the center could also result in a strong outflow.
The spin of the MBH remnant and its orientation are likely to be
well constrained from the preceding {\it LISA} observation of the merger GWs.
As such, the observation of a jet near a recently merged MBH binary
would present an unprecedented opportunity to 
study a MBH-powered jet where the spin of the central engine
is precisely and independently constrained.

We conclude that the observation of the EM accretion afterglow
of a MBH merger is likely to provide a windfall of
empirical constraints on the physics of gas accretion onto MBHs.
Such an observational opportunity would be unprecedented on three points.
First, the ``initial condition'' for the structure of the
evolving flow may be relatively well characterized thanks to
models for circumbinary disks and the theoretical
understanding of the orbital evolution of the binary leading up to the merger.
Second, the accretion flow is likely to evolve on humanly tractable
timescales.
Such an accretion signature will act as a probe of the viscosity in the flow,
in contrast with a steady $\alpha$-disk in which the emission is independent of the $\alpha$ parameter.
Third, the mass, spin and orientation of the central MBH
 will have been independently determined by {\it LISA}.
 The last point is important in terms of emission geometry, energetics,
 and particularly significant if the accretion flow
fuels an outflow or a jet as it fills the central cavity.

In this work, we have used idealized models
with a simple radial power-law prescription for the gas viscosity.
However, in the $\alpha$-viscosity model the viscosity is also a function of
surface density and disk thickness.
In light of the large range of $\Sigma$ and $H/R$ values found in the
regimes of interest in our solutions, it would not be surprising
if a more realistic viscosity prescription led to significantly
different results from those presented here.
Even in our highly idealized calculations, the richness of the physical
problem that these systems represent is readily apparent.
Given advection, radiation-pressure dominance, super-Eddington fluxes,
issues of disk stability and geometry, and rapid viscous evolution,
the circumbinary accretion flows around a merging MBH binary
are likely to produce intriguing observational signatures that would serve as
unprecedented probes of fundamental astrophysical processes.
The interesting possibilities raised by our simple models underscore the need for more detailed
investigations of this interesting class of objects.\\

We thank Zolt\'an Haiman for insightful conversations and constructive comments on the manuscript.
We are grateful to the anonymous referee for comments that helped improve this paper.
This work was supported by NASA ATFP grant NNXO8AH35G.

\newpage
\begin{deluxetable}{ccccccccc}
\tablecolumns{9}
\tablecaption{\label{tab:1}Inner Circumbinary Disk at Decoupling\tablenotemark{a}}
\tablehead{
  \colhead{Variable} &
  \colhead{Factor} &
  \colhead{$\alpha_{-1}$} & 
  \colhead{$S$} &
  \colhead{$\lambda$} &
  \colhead{$M_6$} &
  \colhead{$\beta_{-1},\zeta $} &
  \colhead{$\theta_{0.2}$}
}
\startdata 
$a_{0}/(GM/c^2)$ & 126 & -0.34 & -0.24 & 0.70 & 0.08 & 0.42 & -0.08  \\
$t_{\rm EM}\  (\textrm{yr}$) & 9.2 & -1.36 & -0.98 & 2.80 & 1.32 & 1.7,\ 0.7 &-0.34 \\
$\Sigma_{0} (\g\cm^{-2})$ & $6.2\times 10^{5}$ & -0.68 & 0.51 & -0.60 & 0.16 & -0.15 & -0.17  \\
$T_{0}\ (10^6\textrm{ K})$ & 1.3 & 0.19 & 0.86 & -1.95 & -0.28 & -0.49 & 0.30 \\
$H_{0}/R_{0}$ & 0.17 & 0.76 & 2.43 & -3.80 & -0.12 & -0.95 & 1.19 \\
$P_{\rm rad,0}/P_{\rm gas,0}$ & 430 & 1.67 & 4.25 & -7.35 & -0.04 & -1.84 & 2.17 \\
${\cal Q}_{\rm adv,0}/{\cal Q}_{\rm rad,0}$ & $2.0\times 10^{-2}$ & 1.52 & 4.86 & -7.60 & -0.24 & -1.90 & 2.38 \\
\enddata
\tablenotetext{a}{At decoupling, the variable in column 1 equals the 
factor in column 2 multiplied by the column head parameters
raised to the powers indicated in columns 3--8.
All quantities except for $a_{0}$ are evaluated at the inner edge of the circumbinary disk.
We derive power-law dependencies identical to those found by \citetalias{MP05}.
Note, however, that we derive a somewhat lower temperature than \citetalias{MP05}, and estimate that
the disk is significantly less geometrically thick than they did.
See Appendix \ref{sec:AppA} for detailed calculations.
}
\end{deluxetable}

\newpage

\appendix
\section*{APPENDICES}
\section{Properties of the Circumbinary Disk After Decoupling}
\label{sec:AppA}

We review below the properties of the inner circumbinary disk at the time of decoupling.
Decoupling is defined as the time when $t_{\rm GW}$,
the timescale  on which the binary orbit shrinks,
becomes shorter than $\beta t_{\rm visc}$, the timescale 
on which the central cavity refills due to viscous spreading of the disk.
The procedure is as follows.
First, the decoupling condition $t_{\rm GW}\sim \beta t_{\rm visc}$
relates the gas properties at the inner edge to the orbital parameters of the binary
in terms of the disk midplane temperature.
Second, the midplane temperature is calculated by making simple assumptions
about the thermal structure of the disk.
\citetalias{MP05} performed these calculations for the inner edge of the disk;
we follow their approach closely and reproduce their results
while pointing out a few differences.
We also discuss possible ways to set the radial power $n$ of the prescribed
viscosity in our time-dependent solutions.

Assumptions underlying the model are detailed
in the body of this paper and in \citetalias{MP05}.
Notation for physical quantities is given in our \S\ref{sec:model}.
Fiducial parameter values are $M=10^{6}\Msol M_{6}$, $\zeta\equiv 4q/(1+q)^{2}=1$,
$\alpha=0.1\alpha_{-1}$, $\beta=0.1\beta_{-1}$,  $\lambda=1$, $S=1$, and $\theta_{0.2}=1$.
Unless otherwise specified, the subscript ``0''  refers
to the value at decoupling when applied to binary properties; when applied to
gas properties, it refers to the value at decoupling time and at the inner edge of the disk.

\subsubsection*{A1. Properties at decoupling as functions of midplane temperature}
For circular orbits with semimajor axis $a$, the timescale of
gravitational-wave-driven decay of the binary separation is given by:
\beq
t_{\rm GW}\equiv \frac{a}{da/dt}=\frac{5}{16}\frac{c^{5}}{G^{3}M^{3}}a^{4}\zeta^{-1}.
\eeq
The viscous time is 
\beq
t_{\rm visc}= \frac{2}{3}\frac{R^{2}}{\nu}=\frac{R^{2}\Omega\mu m_{\rm p}}{\alpha k T}
\approx 144 \yr \times \alpha_{-1}^{-1}\lambda^{1/2}T_{6}^{-1}\left(\frac{a}{100 GM/c^{2}}\right)^{1/2},
\eeq
where $T=10^{6}\K\; T_{6}$ is the midplane temperature.

Applying the decoupling condition $t_{\rm GW,0}\sim \beta t_{\rm visc,0}$ gives
the following quantities in terms of the midplane temperature at the inner edge, $T_{0,6}$:
\begin{eqnarray}
a_0&\approx& 136  \frac{GM}{c^{2}}\times
T_{0,6}^{-2/7} \alpha_{-1}^{-2/7}\lambda^{1/7}\left(\beta_{-1}\zeta\right)^{2/7} ,
\label{eq:a0}\\
t_{\rm EM}&=& \frac{3}{4}\beta t_{\rm visc,0}\approx 13 \yr  \times 
T_{0,6}^{-8/7}\alpha_{-1}^{-8/7}\lambda^{4/7}M_{6}\beta_{-1}^{8/7}\zeta^{1/7},
\label{eq:tEM}\\
\nu_{0}
&\approx&2.0\times 10^{17} \cm^{2}\s^{-1}\times T_{0,6}^{4/7}\alpha_{-1}^{4/7}\lambda^{12/7}M_{6}\left(\beta_{-1}\zeta\right)^{3/7},
\label{eq:nu0}\\
\Omega_{0}&\approx&4.5\times 10^{-5}\s^{-1}\times T_{0,6}^{3/7}
\alpha_{-1}^{3/7}\lambda^{-12/7}M_{6}^{-1}\left(\beta_{-1}\zeta\right)^{-3/7}
\label{eq:Omega0}.
\end{eqnarray}

We parametrize the surface density via the arbitrary relation
$\Sigma_{0}\equiv S\dot{M}_{\rm Edd}/(3\pi\nu_{0})$,
where  the parameter $S$ may be interpreted as the product of the outer mass supply rate in Eddington units
and the excess in surface density near the inner edge over the
value expected for a standard Shakura-Sunyaev disk,
which arises from mass accumulation near the binary \citep{IPP99, Chang+09}.
The latter factor depends on tidal interactions and prior accretion history, and
in general is expected to exceed unity.  We obtain 
\beq
\Sigma_{0}\approx 7.3\times 10^{5}\g\cm^{-2}\times T_{0,6}^{-4/7}
\alpha_{-1}^{-4/7}S\lambda^{-12/7}\left(\beta_{-1}\zeta\right)^{-3/7}.
\label{eq:SigDef}
\eeq

The pressure near the edge is dominated by the radiation pressure
$P_{\rm rad}=(4\sigma/3c)T^{4}$, if $T\gtrsim 10^{6}\K$.
We calculate the scale-height-to-radius ratio $H_{0}/R_{0}$
and the radiation-to-gas pressure ratio $P_{\rm rad,0}/P_{\rm gas,0}$ at the disk edge:
\begin{eqnarray}
\frac{H_{0}}{R_{0}}&\approx&\sqrt{\frac{\gamma P_{\rm rad,0}H_{0}}{\Sigma_{0}}}\frac{1}{\Omega_{0}R_{0}}
=\frac{\gamma P_{\rm rad,0}}{\Sigma_{0}\Omega_{0}^{2}R_{0}}\nonumber\\
&\approx& 0.056 \times T_{0}^{4}S^{-1}\lambda^{4}M_{6}\beta_{-1}\zeta,
\label{eq:HoverR}\\
P_{\rm gas,0}&=&\frac{kT_{0} \Sigma_{0}}{\mu m_{\rm p}H_{0}}
= \frac{ kT_{0}\Sigma_{0}^{2} \Omega^{2}}{\gamma \mu m_{\rm p} P_{\rm rad,0}},
\nonumber\\
\frac{P_{\rm rad,0}}{P_{\rm gas,0}}
&=&\frac{\gamma \mu m_{\rm p} P_{\rm rad,0}^{2}}{ kT_{0}\Sigma_{0}^{2} \Omega^{2}}\nonumber\\
&\approx& 57\times T_{0,6}^{51/7}\alpha_{-1}^{2/7}S^{-2}\lambda^{48/7}M_{6}^{2}\left(\beta_{-1}\zeta\right)^{12/7}
\label{eq:PrPg}.
\end{eqnarray}
Above, $\gamma\approx 4/3$ is the relevant adiabatic index,
and $\rho=\Sigma/H$ is the gas density.
(Note that in the $H/R$ profiles plotted in Figures \ref{fig:adv1} and \ref{fig:adv2},
we solve for the scale height $H$ in terms of the total pressure $P_{\rm gas}+P_{\rm rad}$
and do not assume radiation pressure dominance.)

\subsubsection*{A2. Calculating the midplane temperature}
We now turn to calculating the midplane temperature of the inner edge of the disk
at decoupling.
The thermal spectrum of the inner gas disk differs from that of a blackbody
because scattering is a significant source of opacity.
In such a medium, photons travel a shorter effective path before becoming thermalized
and the absorption opacity is effectively enhanced as \citep[e.g.,][]{RL86, Blaes04}: 
\beq
\kappa_{\rm eff,\nu}\approx \kappa_{\rm abs,\nu}\sqrt{1+\frac{\kappa_{\rm es}}{\kappa_{\rm abs,\nu}}}
\label{eq:opaceff}
\eeq
where $\kappa_{\rm abs, \nu}$ and $\kappa_{\rm es}\approx 0.40 \cm^{2}\g^{-1}$ are the
absorption and scattering opacities, respectively.
The second approximation in equation \eqref{eq:opaceff}
is applicable if electron scattering is the dominant source of opacity,
which is generally the case for our problem in the spectral range of interest (UV frequencies and higher).

The disk radiates through a photosphere,
with photons of different energies being thermalized at different depths.
The bottom of the photosphere is defined as the height where the effective optical depth
$\tau_{\rm eff,\nu}\approx \rho_{\rm p}H_{\rm p}\kappa_{\rm eff,\nu}=1$,
where $\rho_{\rm p}$ is the density there and $H_{\rm p}$ is the
scale height of the photosphere.
Characterizing $H_{\rm p}$ as the ratio of the sound speed at the bottom of the photosphere
to the orbital angular speed leads to
\beq
\rho_{\rm p}\approx\frac{3c\Omega^{2}}{4\gamma \sigma  T_{\rm p}^{4}\kappa_{\rm abs,\nu}(\kappa_{\rm abs,\nu}+\kappa_{\rm es})},
\label{eq:rhoTP}
\eeq
where $T_{\rm p}$ is the temperature at the bottom of the photosphere.

The main source of absorption is the bound-free process,
which in general is a complicated function of composition
and photon frequency due to its dependence on the ionization state of the gas.
For consistency and simplicity, we prescribe the absorption opacity in the same way as \citetalias{MP05},
so that the frequency dependence has the same functional form as the free-free absorption opacity, viz.
\beq
\kappa_{\rm abs, \nu}\approx \kappa_{\rm abs, R}~\rho_{\rm p} T_{\rm p}^{-7/2}[187 f_{\nu}(\xi)].
\label{eq:kapTP}
\eeq
Above, $\xi\equiv h\nu/kT_{\rm p}$, $f_{\nu}(\xi)\equiv \xi^{-3}(1-e^{-\xi})$ gives the
frequency dependence, and $\rho_{\rm p}$ and $T_{\rm p}$ are in cgs.
The coefficient $\kappa_{\rm abs, \nu}$ is scaled so that its Rosseland mean
yields the Kramer's opacity scaling at solar metallicity, with 
$\kappa_{\rm abs, R}\approx 1.6 \times 10^{24} \cm^{2}\g^{-1}$ and
the factor in square brackets having a Rosseland mean of unity.
Substituting for the photospheric density using equation \eqref{eq:rhoTP} gives
a cubic equation for the absorption-to-scattering opacity ratio
$K_{\nu}\equiv \kappa_{\rm abs, \nu}/\kappa_{\rm es}$
 at the bottom of the photosphere:
\beq
\kappa_{\rm abs,\nu}^{2}\left(1+\frac{\kappa_{\rm abs,\nu}}{\kappa_{\rm es}}\right)
\approx
140 \frac{c\kappa_{\rm abs,R}}{\gamma \sigma \kappa_{\rm es}}\Omega^{2}T_{\rm p}^{-15/2}f_{\nu}(\xi).
\label{eq:kapcube}
\eeq
We find that in most cases, the vast majority of the disk emission
is emitted from in frequencies and regions that are scattering dominated,
i.e. $\kappa_{\rm abs,\nu}/\kappa_{\rm es}\ll 1$.
In this limit, \eqref{eq:kapcube} may be solved trivially to obtain
$\kappa_{\rm abs, \nu}\approx \kappa_{\rm abs,*} f_{\nu}^{1/2}$,
where we have defined for convenience the frequency-independent quantity
$\kappa_{\rm abs,*}\approx 4.7\times 10^{20 }\cm^{2}\g^{-1}\times 
(\Omega \s) (T_{\rm p}/\K)^{-15/4}$.
However, in some of our solutions there is significant emission from
regions with $\kappa_{\rm abs,\nu}/\kappa_{\rm es} \gtrsim 1$.
All of the spectra and radial disk profiles in our figures were computed
by solving \eqref{eq:kapcube} generally.

The angle-integrated emergent spectral flux \citep[e.g.,][]{Blaes04} is
\beq
F_{\nu}= \pi \frac{2\epsilon_{\nu}^{1/2}}{1+\epsilon_{\nu}^{1/2}}B_{\nu},
\label{eq:modF}
\eeq
where $B_{\nu}$ is the Planck function and 
$\epsilon_{\nu}\equiv\kappa_{\rm abs,\nu}/(\kappa_{\rm abs,\nu}+\kappa_{\rm es})=(1+K_{\nu}^{-1})^{-1}\le 1$ is the
frequency-dependent ratio of the absorption opacity to the total opacity.
Both $\epsilon_{\nu}$ and $B_{\nu}$ are evaluated at the bottom of the photosphere.
Each face of the disk contributes half of the total flux $F$.
Integrating equation \eqref{eq:modF}, we obtain:
\beq
\frac{F}{2}=\int_{0}^{\infty}F_{\nu}\;d\nu =
\Xi\sigma T_{\rm p}^{4},
\label{eq:bolF}
\eeq
where $\Xi$ is the deviation of the flux from blackbody,
\beq
\Xi\equiv
\frac{15}{\pi^{4}}
\int_{0}^{\infty}\frac{2\epsilon_{\nu}^{1/2}}{1+\epsilon_{\nu}^{1/2}(\xi)}
\frac{e^{-\xi}d\xi}{f_{\nu}(\xi)}.
\label{eq:Xi}
\eeq
$\Xi$ may be expressed as a function of the frequency-independent quantity
$\epsilon_{*}(T_{\rm p},\Omega)\equiv 
\kappa_{\rm abs,*}/(\kappa_{\rm abs,*}+\kappa_{\rm es})\le 1.$
In the limit $\epsilon_{\nu}^{1/2}\ll 1$, $\Xi \approx 30 \pi^{-4}\epsilon_{*}^{1/2}\int_{0}^{\infty}f_{\nu}^{~-3/4}e^{-\xi}\;d\xi\approx 0.873\epsilon_{*}^{1/2}.$
In the opposite limit $\epsilon_{*}\approx 1$ (i.e. very close to blackbody),
$\Xi\approx 1/[1+(\epsilon_{*}^{-1}-1)^{2/3}]$.
We find that the brightest regions of our disk generally span the
entire range $0\le \epsilon_{*} \le 1$.
We find a useful fitting formula for $\Xi$ to be
\beq
\Xi\approx \frac{0.873\epsilon_{*}^{-1/6}}{1-0.127\epsilon_{*}^{5/6}}
\frac{1}{1+\left(\epsilon_{*}^{-1}-1\right)^{2/3}}.
\eeq
The above prescription evaluates $\Xi$ accurately to within $\sim 1\%$
for the entire range of $\epsilon_{*}$
while reproducing the asymptotic behavior at the extreme limits $\epsilon_{*}\ll 1$ and $\epsilon_{*}\approx 1$.
We also find that a much simpler formula,
$\Xi\approx (4/5) \epsilon_{*}^{1/2}$,
calculates $\Xi$ accurately to within $10\%$ for $0\le \epsilon\ltsim 0.9$.

The optical depth between the bottom
of the photosphere and the midplane is dominated by electron scattering,
and thus has negligible frequency dependence.
Provided that this region is optically thick and can be described
by a one-zone formalism, the midplane and photosphere temperatures
are related by
 $T_{\rm p}^{4} =(4/3)T^{4}/\tau$, where $\tau$ is the optical thickness
 between the two heights.
Following \citetalias{MP05}, we write
$\tau=\theta\kappa_{\rm es}\Sigma$, where $\theta=\theta_{0.2}/0.2\le1$
is a porosity correction factor.
Ignoring horizontal advection, we equate $F$ with the power viscously dissipated per unit area in the disk:
\beq
F=2\times \Xi(\epsilon_{*})\frac{4\sigma T^{4}}{3\tau}\approx  \frac{32\sigma T^{4}\epsilon_{*}^{1/2}}{15\tau}\sim \frac{9}{4}\nu\Sigma\Omega^{2}.
\label{eq:fluxeq}
\eeq

We now solve for $T_{0,6}$ by substituting
equations \eqref{eq:nu0}, \eqref{eq:Omega0} and \eqref{eq:SigDef} into equation \eqref{eq:fluxeq}.
A useful intermediate result is
\beq
\epsilon_{*}^{1/2}\approx\sqrt{\frac{\kappa_{\rm abs,*}(\Omega,T)}{\kappa_{\rm es}}}
\approx 0.19T_{0,6}^{-27/14}\times 
\alpha_{-1}^{-3/56}S^{15/32}\lambda^{-93/56}M_{6}^{-1/2}\left(\beta_{-1}\zeta\right)^{-93/224}\theta_{0.2}^{15/32},
\eeq
which gives $T_{0,6}\propto( \nu_{0}\Sigma_{0}\Omega_{0}^{2}\epsilon_{*}^{-1/2}\tau_{0})^{14/25}$.
That is, the midplane temperature scales with the flux scaling to the $+14/25$th power, i.e.
more sensitively than in the standard black body relation of $T\propto F_{\rm bb}^{1/4}$.
We finally arrive at
\beq
T_{0,6}\approx 1.3 \times 
\alpha_{-1}^{19/100}S^{343/400}\lambda^{-39/20}M_{6}^{-7/25}\left(\beta_{-1}\zeta\right)^{-39/80}\theta_{0.2}^{119/400}.
\eeq

We derive an inner-edge temperature that is somewhat lower than the
value calculated by \citetalias{MP05}, primarily because
their prescription for the emitted flux is lower than ours by a factor of two.
(Compare their expression for $F_{\nu}$ with our equation \eqref{eq:modF}.
See also our discussion that follows equation \eqref{eq:modF1}.)
As noted above, the midplane temperature is rather sensitive to the flux.
We calculate that \citetalias{MP05} underestimated the flux through the omission of the
factor of two, and slightly overestimated the frequency integral.
Overall, we calculate a bolometric flux $60\%$ higher than they did,
and our disk temperature is lower by a factor $1.6^{14/25}\approx 1.3$.  
We derive identical power-law dependencies on the system parameters as they did.

The lower temperature suggests a slightly longer timescale for
the nominal onset of the X-ray afterglow after merger,
$t_{\rm EM}\sim (3/4)\beta t_{\rm visc,0}\propto T_{0}^{-8/7}$
$\sim 9\times M_{6}^{1.3}\yr$ for fiducial parameters.
We also calculate lower values
for the quantities $H_{0}/R_{0}$ and $P_{\rm rad,0}/P_{\rm gas,0}$,
which are highly sensitive to the disk temperature
(equations [\ref{eq:HoverR}] and [\ref{eq:PrPg}]) and thus significantly
reduced when the factor of two correction to the flux is included.
We summarize our results below, and in our Table \ref{tab:1}.
\begin{eqnarray}
a_{0}&\approx&126 \frac{GM}{c^{2}}\times
\alpha_{-1}^{-17/50}S^{-49/200}\lambda^{7/10}M_{6}^{2/25}\left(\beta_{-1}\zeta\right)^{17/40}\theta_{0.2}^{-17/200},
\\
\Omega_{0}&\approx&5.1\times 10^{-5}\s^{-1} \times
\alpha_{-1}^{51/100}S^{147/400}\lambda^{-51/20}M_{6}^{-28/25}\left(\beta_{-1}\zeta\right)^{-51/80}\theta_{0.2}^{51/400},
\\
t_{\rm EM}&\approx& 9.2\yr \times
\alpha_{-1}^{-34/25}S^{-49/50}\lambda^{14/5}M_{6}^{33/25}\beta_{-1}^{17/10}\zeta^{7/10}\theta_{0.2}^{-17/50},
\\
\Sigma_{0}&\approx& 6.2\times 10^{5} \g \cm^{-2} \times 
\alpha_{-1}^{-17/25}S^{51/100}\lambda^{-3/5}M_{6}^{4/25}\left(\beta_{-1}\zeta\right)^{-3/20}\theta_{0.2}^{-17/100},
\\
\frac{H_{0}}{R_{0}}&\approx& 0.17\times 
\alpha_{-1}^{19/25}S^{243/100}\lambda^{-19/5}M_{6}^{-3/25}\left(\beta_{-1}\zeta\right)^{-19/20}\theta_{0.2}^{119/100},
\\
\frac{P_{\rm rad,0}}{P_{\rm gas,0}}&\approx& 430\times 
\alpha_{-1}^{167/100}S^{1699/400}\lambda^{-147/20}M_{6}^{-1/25}\left(\beta_{-1}\zeta\right)^{-147/80}\theta_{0.2}^{867/400},\\
\epsilon_{*}^{1/2}&\approx& 0.11\times 
\alpha_{-1}^{-21/50}S^{-237/200}\lambda^{21/10}M_{6}^{1/25}\left(\beta_{-1}\zeta\right)^{21/40}\theta_{0.2}^{-21/200}.
\end{eqnarray}
We calculate a somewhat shorter scale height for the disk than \citetalias{MP05},
who had derived $H_{0}/R_{0}\sim 0.46$
for fiducial parameters.
We show in \S \ref{sec:adv} that the advected flux at decoupling is not likely to be significant
in the circumbinary disk (for $S=1$).

\subsubsection*{A3. Prescribing a value for the viscosity power-law index $n$}
\label{sec:nval}

In our model we prescribe
a simple radial power-law $\nu\propto R^{n}$ for the viscosity.
In the $\alpha_{\rm gas}$-disk model, however, $\nu$
has a physical definition with
$\nu\propto \alpha P_{\rm gas}/(\rho\Omega)\propto T/\Omega$.
This suggests that a reasonable value for $n$ should satisfy the relation
\beq
n\approx \frac{d\ln T}{d\ln r}+3/2.
\label{eq:nandT}
\eeq
Below, we apply the above relationship between $\nu$ and $T$ to
evaluate an appropriate value for our viscosity
power-law index $n$.

For fiducial parameters, near the disk edge
at decoupling and inside this radius thereafter,
$\Xi\approx (4/5)\epsilon_{*}^{1/2}$ and 
$\epsilon_{*}\approx \kappa_{\rm abs,*}/\kappa_{\rm es}\propto \Omega T_{\rm p}^{-15/4}$.
Applying these approximations to equation \eqref{eq:fluxeq} leads to
$T_{\rm p}^{17/8}\propto \nu\Sigma\Omega^{3/2}$.
Since $T\propto \tau^{1/4}T_{\rm p}\propto \Sigma^{1/4}T_{\rm p}$, we obtain
\beq
T(r,t)\propto\left[\frac{\Sigma(r,t)}{\Sigma_{0}}\right]^{49/68}r^{(8n-18)/17}.
\label{eq:Tofr}
\eeq

One way of characterizing the accretion flow
is through the power-law index $m\equiv d\ln (\nu\Sigma)/d\ln R$.
Writing $d\ln \Sigma/d\ln R =m-n$ and substituting this
into equations \eqref{eq:nandT} and \eqref{eq:Tofr} leads to the solution $n\approx (30+49m)/85$.
If the disk just outside the inner edge behaves like a
steady-state accretion solution with $m\approx 0$,
then equation \eqref{eq:nandT} suggests that in this
region a reasonable value of $n$ is
$6/17\approx 0.35$.
If, however, the gas in this region behaves more like
a decretion profile with $m\approx -1/2$, then
$n=11/170\approx 0.065$ would be more appropriate.

The disk surface density profile outside the edge at decoupling, however,
is not particularly relevant to the viscous refilling rate
of the central cavity.
Rather, the pertinent value of $n$ is that where the torque gradient
is highest, where the most rapid evolution occurs, i.e. near the inner
edge of the disk.
However, in this region  $d\ln(T/\Omega)/d\ln R>0$ is a
rapidly changing function of time and radius.
Thus, there is no single value for $n$
that can fully describe the viscous evolution of an $\alpha$-disk.
At late times ($t\gtrsim t_{\rm visc}(R)$ after the merger), however,
the inner disk evolves toward a standard quasi-steady accretion solution with
$\nu\Sigma\propto (1-\sqrt{R_{*}/R})$, where $R_{*}$ is the inner boundary
radius imposed by the MBH remnant.  In this limit,
equation \eqref{eq:nandT} leads to 
$n\approx 6/17+49/(170/\sqrt{R/R_{*}-1})$, which
evaluates to $n\approx 0.4$ at radii $R\ltsim R_{0}$
for $1\le R_{*}c^{2}/GM\le 6$.
We thus select $n=0.4$ as our fiducial viscosity power-law index,
as it is consistent with the asymptotic evolution of the inner disk at late times.

Upon experimenting with several values for $n$,
we have found that the choice of $n$ does not qualitatively
affect the main conclusions of our study,
as long as $n\ltsim 1$.
The viscous evolution is driven by the single timescale
$\tau\propto t_{\rm visc}/(1-n/2)^{2}$, and
the bolometric and monochromatic
luminosities also have weak direct dependencies on $n$.

\section{Green's Function for the Viscous Evolution of the Disk Surface Density}
\label{sec:AppB}

Equation \eqref{eq:diff}, a second-order diffusion equation, is linear if $\nu$ does not
depend on $\Sigma$.  Below, we follow the formalism of \citeauthor{LP74} (\citeyear{LP74}; see also \citealt{Ogilvie05})
to derive a Green's function solution for the viscous evolution.

Suppose the solution has the exponentially
decaying form $\Sigma(R,t)=\exp(-\Lambda t)R^{p}\varsigma(R)$,
where $\varsigma(R)$ is an arbitrary function of radius and $p$ is an arbitrary real number.
With the additional assumption that $\nu$ behaves as a power-law,
$\nu=\nu_{0}(R/R_{0})^{n}$, equation \eqref{eq:diff} may be rewritten as a modified
Bessel equation:
\beq
R^{2}\frac{\dd^{2}\varsigma}{\dd R^{2}}+\left(2p+2n+\frac{3}{2}\right)R\frac{\dd \varsigma}{\dd R}
+\left[\left(p+n\right)\left(\frac{\Lambda}{3\nu_{0}}R^{2-n}+p+n+\frac{1}{2}\right)\right]\varsigma=0.
\label{eq:Bessel}
\eeq
Selecting $p=n-1/4$ for convenience and
making the substitution $\Lambda=3\nu_{0}k^{2}R_{0}^{-n}$,
the equation above has a set of solutions
\beq
\varsigma_{k}(R)=R^{-2n}\left[A(k) J_{1/(4-2n)}\left(\frac{kR^{1-n/2}}{1-n/2}\right)
+B(k)Y_{1/(4-2n)}\left(\frac{kR^{1-n/2}}{1-n/2}\right)\right],
\label{eq:elemsol}
\eeq
where $n< 2$, 
$J$ and $Y$ are the ordinary Bessel functions of the first and second kind, respectively,
$k>0$ is their mode, and $A(k)$ and $B(k)$ are arbitrary coefficients for each mode.

We note here that even with a simple prescription for $\nu$,
the general problem of the viscous evolution of a circumbinary disk
around a binary is not easily treated analytically.
For a case where the Green's function is particularly simple,
\cite{Pringle91} showed that one may account for the binary torques
by applying a zero-torque boundary condition $\dd(\nu\Sigma R^{1/2})/\dd R=0$
at the gap opening radius.
For binaries of interest in this paper,
the gap-opening radius itself is a function of time, which greatly adds to the
``extreme algebraic complexity'' \citep{Pringle91} that is generally involved with
applying such a boundary condition at nonzero radius.
Our problem, however, is greatly simplified by the decoupling condition,
which allows us to approximate the early evolution as if the circumbinary disk
experiences no torques from the binary.
We thus proceed as if the
potential is due to a single central point mass, and approximate
the gas orbits as being circular and Keplerian.
We discuss the justifications and the modest evolutionary
effects associated with this model simplification in \S \ref{sec:viscev}.

The boundary condition of interest, then, is that the torques vanish
at the center.  (At late times, it will be necessary to consider
the inner boundary condition imposed at finite radius by the central
MBH remnant.  We shall address this point shortly.)
Because $Y_{1/(4-2n)}(ky)$ diverges as $y\rightarrow 0$,
our physical solution requires all $B(k)$ to vanish.
The general solution is a sum of all possible modes over all $k\ge 0$ weighted by a set
of coefficients $A(k)$, viz.:
\beq
\Sigma(R,t)=R^{-n-1/4}\int_{0}^{\infty}A(k)J_{1/(4-2n)}(ky)
\exp\left[-3\nu_{0}R_{0}^{-n}k^{2}t\right]~dk
,
\label{eq:sig1}
\eeq
where we have defined $y\equiv R^{1-n/2}/(1-n/2)$ for convenience.

The function $A(k)$ can be evaluated through the use of Hankel transforms.
For a given order of the Bessel function $\ell$, Hankel transform pairs satisfy:
\begin{eqnarray}
\Phi(R)&=&\int_{0}^{\infty}\phi(k)J_{\ell}(kR)~k\;dk\\
\phi(k)&=&\int_{0}^{\infty}\Phi(R)J_{\ell}(kR)~R\;dR
\end{eqnarray}
For our problem, we may construct the Hankel transform pair
\begin{eqnarray*}
\Phi(y)&=&R^{n+1/4}\Sigma(y,0)=
\int_{0}^{\infty}\left[k^{-1}A(k)\right]J_{1/(4-2n)}(ky) ~k\; dk\\
\phi(k)&=&k^{-1}A(k)=\int_{0}^{\infty}\left[R^{\prime n+1/4}
\Sigma(y^{\prime},0)\right]J_{1/(4-2n)}(ky^{\prime})~y^{\prime}\; dy^{\prime},
\end{eqnarray*}
from which we obtain
\begin{eqnarray}
A(k)&=&\int_{0}^{\infty}R^{\prime n+1/4}
\Sigma(y^{\prime},0)~J_{1/(4-2n)}(ky^{\prime})\;k\; y^{\prime}~dy^{\prime}\nonumber\\
&=&\left(1-\frac{n}{2}\right)^{-1}\int_{0}^{\infty}
\Sigma(y^{\prime},0)~J_{1/(4-2n)}(ky^{\prime})\;k\; R^{\prime 5/4+n/2} ~dR^{\prime}
\label{eq:sig2}
\end{eqnarray}

Combining equations \eqref{eq:sig1} and (\ref{eq:sig2}),
we may write the solution $\Sigma(y,t)$ as an integral function of
the initial condition $\Sigma(R^{\prime},0)$,
\begin{eqnarray}
\Sigma(R,t)&=&\left(1-\frac{n}{2}\right)^{-1}R^{-n-1/4}\int_{0}^{\infty}R^{\prime 5/4+n/2}\int_{0}^{\infty}
\Sigma(R^{\prime},t=0)\,\exp\left(-3\nu_{0}R_{0}^{-n}k^{2}t\right)\nonumber\\
&&\qquad \times 
J_{1/(4-2n)}(ky^{\prime})J_{1/(4-2n)}(ky) ~ k\;dk\;dR^{\prime}
\nonumber\\
&\equiv&\int_{0}^{\infty}G_{0}(R,R^{\prime}, t)\Sigma(R^{\prime},t=0)\frac{dR^{\prime}}{R_{0}},
\label{eq:solution}
\end{eqnarray}
with the Green's function
\begin{eqnarray}
G_{0}
&=&\left(1-\frac{n}{2}\right)^{-1}R^{-1/4-n}R^{\prime 5/4}R_{0}
\int_{0}^{\infty}J_{1/(4-2n)}(ky^{\prime})\; J_{1/(4-2n)}(ky)
\exp\left(-3\nu_{0}R_{0}^{-n}k^{2}t\right)
\; k\;dk\nonumber\\
&=&\frac{R^{-1/4-n}R^{\prime 5/4}R_{0}^{1+n}}{6(1-n/2)\nu_{0} t}
I_{1/(4-2n)}\left[\frac{R^{1-n/2}R^{\prime 1-n/2}R_{0}^{n}}{6(1-n/2)^{2}\nu_{0} t}\right]
\exp\left[-\frac{\left(R^{2-n}+R^{\prime 2-n}+\right)R_{0}^{n}}{12(1-n/2)^{2}\nu_{0} t}\right]\nonumber\\
&=&\frac{2-n}{\tau}r^{-1/4-n}r^{\prime 5/4}
I_{1/(4-2n)}\left(\frac{2r^{1-n/2}r^{\prime 1-n/2}}{\tau}\right)
\exp\left(-\frac{r^{2-n}+r^{\prime 2-n}}{\tau}\right),
\label{eq:GreenApp}
\end{eqnarray}
where $I_{\ell}$ is the modified Bessel function of the first kind and order $\ell$, and
we have substituted the dimensionless variables
$r\equiv R/R_{0}$ and $\tau\equiv 8(1-n/2)^{2}t/t_{\rm visc,0}$.

In deriving the Green's function above,  we have applied the zero-torque boundary
condition at the origin, instead of at some finite boundary radius $R_{*}$.
At late times and small radii, $\tau\gg  r^{2-n}$,
the solutions obtained using equation \eqref{eq:GreenApp}
have the radial dependence $\dd \ln (\nu\Sigma)/\dd \ln R\approx (n-2)r^{2-n}/\tau$,
viz. they converge to a quasi-steady profile with $\nu\Sigma\propto \Sigma r^{n}$ constant in radius.
Thus, $\Sigma$ diverges at the center at late times if $n$ is positive.

A physically plausible model should account for the boundary condition
imposed by the MBH remnant at late times.
Below, we approximate the Green's function for the case where
the zero-torque boundary condition is applied at a finite
boundary radius $R_{*}>0$.
In this case the boundary condition determines the relationship between
the $A(k)$ and $B(k)$ coefficients in equation \eqref{eq:elemsol}, viz:
\beq
\varsigma_{k}(R)=R^{-2n}A(k)
\left[J_{1/(4-2n)}(ky)-
\frac{J_{1/(4-2n)}(ky_{*})}{Y_{1/(4-2n)}(ky_{*})}
Y_{1/(4-2n)}(ky)
\right],
\eeq
where $y_{*}=R_{*}^{1-n/2}/(1-n/2)$.

Our new Green's function is then
\begin{eqnarray}
G&=&R^{-1/4-n}R^{\prime 5/4+n/2}
\int_{0}^{\infty}
J_{1/(4-2n)}(ky)\;J_{1/(4-2n)}(ky^{\prime})
\exp\left[-\frac{R_{0}^{2-n}k^{2}}{4\left(1-n/2\right)^{2}}\tau\right]\nonumber\\
&&\qquad \times
\left[1-\frac{J_{1/(4-2n)}(ky_{*})}{Y_{1/(4-2n)}(ky_{*})}\;
\frac{Y_{1/(4-2n)}(ky)}{J_{1/(4-2n)}(ky)}\right]\;k\;dk.
\label{eq:GreenBC1}
\end{eqnarray}
The integrand above is greatest where $k$ is small,
and decays rapidly as $k$ increases beyond $k^{2}\sim 4R_{0}^{n-2}\tau^{-1}$.
The boundary condition, however, has a significant effect on the
viscous evolution only where $\tau\gg r^{2-n}=y^{2}R_{0}^{n-2}$
and $y\gtrsim y_{*}$.  Thus, the
boundary effect is appreciable only where $k^{2}\ll 4y_{*}^{-2}\gtrsim 4y^{-2}$.
In the limit of small argument $J_{\ell}(x)/Y_{\ell}(x)\propto x^{2m}$,
and we obtain
\beq
\frac{J_{1/(4-2n)}(ky_{*})}{Y_{1/(4-2n)}(ky_{*})}\;
\frac{Y_{1/(4-2n)}(ky)}{J_{1/(4-2n)}(ky)}
\approx \left(\frac{y_{*}}{y}\right)^{1/(2-n)}=\sqrt{\frac{R_{*}}{R}}.
\eeq
To leading order, then, at late times and small radii
the factor in square brackets in equation \eqref{eq:GreenBC1}
evaluates to $1-\sqrt{R_{*}/R}$.
This factor may be taken outside of the integral,
and gives the asymptotic behavior $\nu\Sigma\propto 1-\sqrt{R_{*}/R}$,
which is precisely the standard solution for a steady thin accretion disk 
with a no-torque inner boundary at $R_{*}$ \citep[e.g.,][]{FKR02}.

We do not evaluate the exact Green's function solution
at early times or at large radii, where the effect of the boundary
condition is minimal.  
Instead, we approximate the function by ensuring
the correct boundary behavior at late times and small radii,
as suggested by \cite{LP74}.
Accounting for the fact that the boundary at $R_{*}$ exists only 
after the time of merger, which evaluates via the decoupling condition
to $\tau_{\rm merge}=2\beta(1-n/2)^{2}$, we approximate
our new Green's function as follows:
\beq
G(r,r^{\prime},\tau)=\left\{1-\sqrt{\frac{r_{*}}{r}}
\exp\left[-\frac{\left(r-r_{*}\right)^{2-n}}{{\mathcal R}(\tau-\tau_{\rm merge})}\right] \right\}
G_{0}(r,r^{\prime},\tau),
\label{eq:GreenBC}
\eeq
where $G_{0}$ is the Green's function given in equation \eqref{eq:GreenApp}
for the case $R_{*}=0$, $r_{*}\equiv R_{*}/R_{0}$,
and ${\mathcal R}$ is the ramp function,
defined as ${\mathcal R}(x\le 0)=0$ and ${\mathcal R}(x>0)=x$.

\bibliographystyle{apj}
\bibliography{TM09}
\end{document}

%% file: TM09_preamble.tex
\newcommand\lsim{\mathrel{\rlap{\lower4pt\hbox{\hskip1pt$\sim$}}
        \raise1pt\hbox{$<$}}}
\newcommand\gsim{\mathrel{\rlap{\lower4pt\hbox{\hskip1pt$\sim$}}
        \raise1pt\hbox{$>$}}}

\newcommand{\K}{\mathrm{K}}

\newcommand{\g}{\mathrm{g}}

\newcommand{\dd}{\partial}

\newcommand{\cm}{\;\mathrm{cm}}

\newcommand{\s}{\;\mathrm{s}}

\newcommand{\km}{\;\mathrm{km}}
\newcommand{\yr}{\;\mathrm{yr}}

\newcommand{\eV}{\; \mathrm{eV}}
\newcommand{\keV}{\; \mathrm{keV}}
\newcommand{\Msol}{M_{\odot}}

\defcitealias{MP05}{MP05}
    \def\dd{\partial}
    \def\tilde{\widetilde}

    \def\beq{\begin{equation} }
    \def\eeq{\end{equation} }
    \def\spose#1{\hbox to 0pt{#1\hss}}
    \def\ltsim{\mathrel{\spose{\lower.5ex\hbox{$\mathchar"218$}}
     \raise.4ex\hbox{$\mathchar"13C$}}}

\def\tilde{\widetilde}
\def\spose#1{\hbox to 0pt{#1\hss}}
\def\lta{\mathrel{\spose{\lower 3pt\hbox{$\mathchar"218$}}
        \raise 2.0pt\hbox{$\mathchar"13C$}}}
\def\gta{\mathrel{\spose{\lower 3pt\hbox{$\mathchar"218$}}
        \raise 2.0pt\hbox{$\mathchar"13E$}}}
\numberwithin{equation}{section}